\documentclass[lettersize,journal]{IEEEtran}
\usepackage{amsmath,amsfonts}

\usepackage{array}

\usepackage{textcomp}
\usepackage{stfloats}
\usepackage{url}
\usepackage{verbatim}
\usepackage{graphicx}
\usepackage{cite}

\usepackage[whole]{bxcjkjatype}
\usepackage{svg}
\usepackage{comment}
\usepackage{booktabs}
\usepackage{bm}
\usepackage{float}
\usepackage[hang,small,bf]{caption}
\usepackage[subrefformat=parens]{subcaption}
\usepackage[ruled,vlined]{algorithm2e}

\usepackage{enumerate}

\usepackage{multirow}

\newcounter{app}
\renewcommand{\theapp}{\Alph{app}}

\begin{document}

\title{ShibuyaSocial: Multi-scale Model of Pedestrian Flows\\ in Scramble Crossing}

\author{Akihiro Sakurai$^{1}$, Naoya Kajio$^{1}$, and Ko Yamamoto$^{1}$
\thanks{*This work is supported by JST PRESTO Grant Number JPMJPR226A}
\thanks{$^{1}$Akihiko Sakurai, Naoya Kajio and Ko Yamamoto are with Department of Mechano-informatics, The University of Tokyo, 7-3-1 Hongo, Bunkyo-ku, Tokyo, 113-8656, Japan.
        {\tt\small sakurai@ynl.t.u-tokyo.ac.jp}}%
}

\maketitle

\begin{abstract}

This paper presents a learning-based model of pedestrian flows that integrates multi scale behaviors such as global route selection and local collision avoidance in urban spaces, particularly focusing on pedestrian movements at Shibuya scramble crossing.
Since too much congestion of pedestrian flows can cause serious accidents, mathematically modeling and predicting pedestrian behaviors is important for preventing such accidents and providing a safe and comfortable environment.
Although numerous studies have investigated learning-based modeling methods, most of them focus only on the local behavior of pedestrians, such as collision avoidance with neighbors and environmental objects.
In an actual environment, pedestrian behavior involves more complicated decision making including global route selection.
Moreover, a state transition from stopping to walking at a traffic light should be considered simultaneously.
In this study, the proposed model integrates local behaviors with global route selection, using an Attention mechanism to ensure consistent global and local behavior predictions.
We recorded video data of pedestrians at Shibuya scramble crossing and trained the proposed model using pedestrian walking trajectory data obtained from the video.
Simulations of pedestrian behaviors based on the trained model qualitatively and quantitatively validated that the proposed model can appropriately predict pedestrian behaviors.

\end{abstract}



\section{INTRODUCTION}
Modeling of pedestrian flows is an important technology in to predicting crowded situations.
Especially at a festival, concert or some event, pedestrian flows can easily overcrowd or get stuck due to inappropriate guidance or building layout, which  sometimes causes a serious crowd-crush accident \cite{kaneda2005developing,helbing2012crowddisasterssystemicfailures,liang2024unraveling}.
In 2001, the Akashi pedestrian bridge accident in Japan \cite{kaneda2005developing} was a crowd crush accident.
This accident occurred at a fireworks festival because of inappropriate guidance of visitors and resulted in 11 deaths and 247 injured \cite{akashi2002_fireworks_accident_report}.
Most recently, a tragic accident occurred at a Halloween event in Seoul, Korea, 2022 \cite{liang2024unraveling}.
In this accidents, approximately 130,000 people visited the area of the Halloween event, which was just after the COVID-19 pandemic.
Those people were stuck in a narrow street, which caused crowd crush and resulted in 159 deaths and 196 injured \cite{son2025158}.
Prevent such serious accidents necessitates mathematically modeling pedestrian flows and quantitatively simulating and predicting congestions in advance.


Many studies exist for modeling pedestrian flows in the fields of computer graphics, robotics, and urban design.
The social force model \cite{SFModel1995}, which is a representative example of a rule-based model incorporating pedestrian behavior, was extended using the collision avoidance methods studied in the field of robotics \cite{Fiorini1998, Berg2008, Gonon2022RAL}.
Similar rule-based models include the cellular automata \cite{Asano2006} and continuum models \cite{Henderson1974,Yamamoto2013,Yamamoto2021SWARM}.
These studies demonstrate through simple rules self-organizing phenomena such as lane formation in counter flows \cite{Murakami, helbing1998self} and diagonal stripe pattern in crossing flow \cite{cividini2013crossing}.
Data-driven approaches are also employed for modeling pedestrian flows in many studies, employing the advancements in machine learning.
Modeling methods using long short-term memory (LSTM) \cite{Alahi2016, Xue2018}, Attention mechanism \cite{Vemula2018}, and reinforcement learning \cite{chandra2024multi} can predict more complicated behavior of pedestrians considering the surrounding environmental information.
These pedestrian flow models can be applied to the evaluation of congestion in buildings \cite{hoogendoorn2004}, pedestrian guidance \cite{Yamamoto2013}, analysis of crowd crush accidents \cite{helbing2012crowddisasterssystemicfailures}, 
and virtual reality (VR) simulators \cite{berton2020crowd,sakurai2023vr,ishibashi2025haptic}.
\begin{figure}[t]
  \begin{center}
  \includegraphics[width=0.8\hsize]{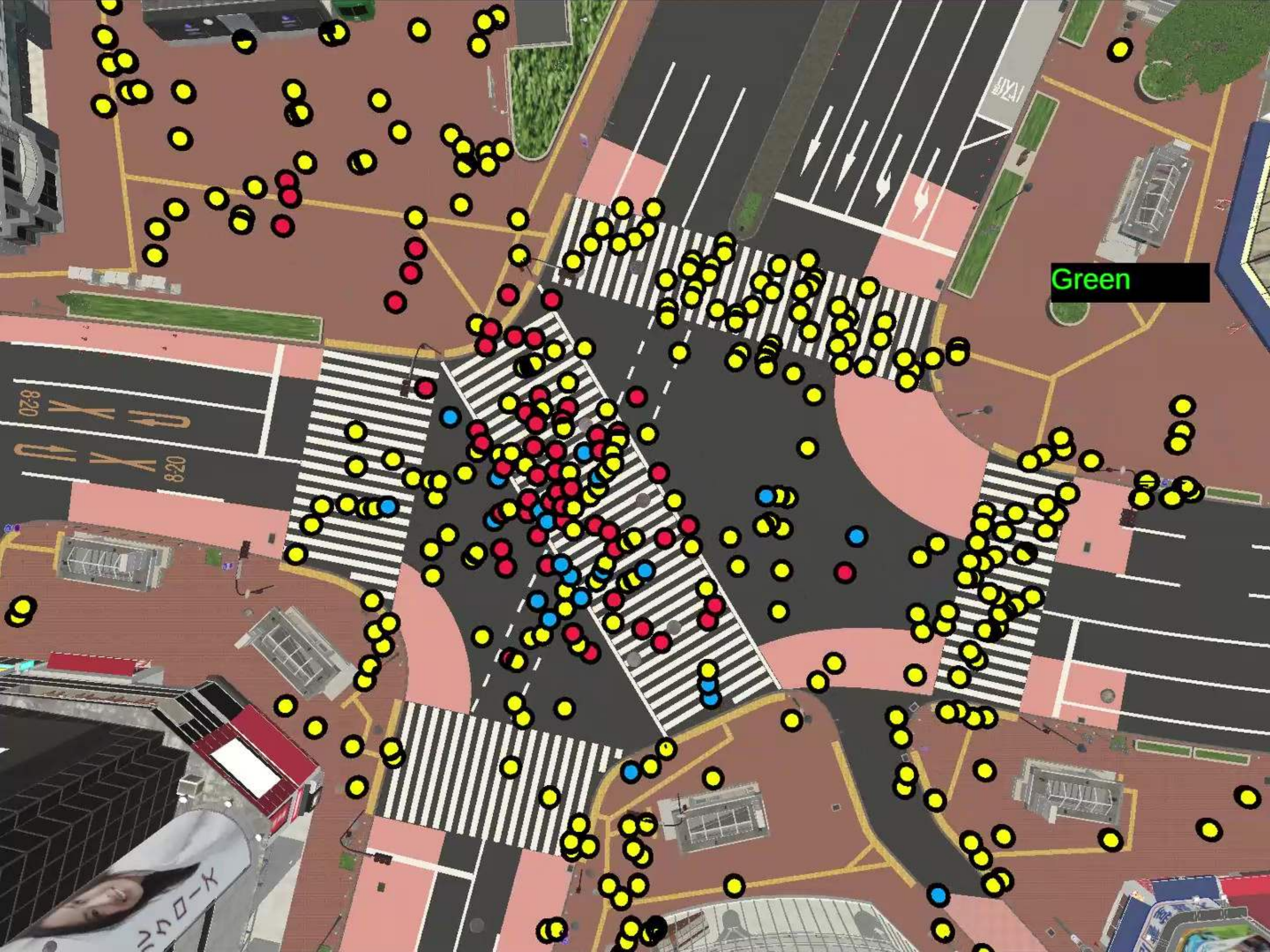}
  \caption{Simulation of two pedestrian flows at Shibuya scramble crossing on a normal day. Red and blue points indicate pedestrian positions in Flows 1 and 2, respectively, whereas yellow points indicate positions of other flows obtained from the measurement data.}
  \label{fig:SHIBUYA_NORMAL_SIMULATION}
  \end{center}
\end{figure}

Most of the aforementioned studies focus on local behaviors at the scale of several tens of cm, such as collision avoidance among pedestrians.
However, pedestrian behaviors in real life depend on more complicated decision-making processes using various types of information.
For example, modeling of route selection behavior at the scale of several tens of meters \cite{Pettre2005,Pelechano2007,Guy2010} is important when crossing a crowded crosswalk.
Moreover, discrete events such as stopping and restart walking should be considered at pedestrian traffic lights.
There are a few models that simultaneously consider route selection and collision avoidance \cite{tran2021goal,wang2022stepwise} on a scale of approximately 10 m.
However, they do not consider pedestrian movement at branching paths and do not account for discrete events such as pedestrian traffic lights.
We proposed a multi scale model that integrated local collision avoidance and global route selection based on the surrounding congestion state and environmental information \cite{sakurai2024learning,sakurai2024spatio}.
However, that study focused on a basic investigation using dummy data generated from the velocity vector field \cite{Yamamoto2013} for training.
Unnatural behaviors such as ignoring a traffic light status were observed when simulating pedestrian behavior based on the trained model.

In this study, the multi scale model proposed in \cite{sakurai2024spatio} with an Attention mechanism \cite{vaswani2017attention} is used to keep the global and local behavior predictions consistent. 
After recording a video of pedestrians at Shibuya scramble crossing, we trained the proposed model using the pedestrian walking trajectory data obtained from the video.
Simulations of pedestrian behaviors based on the trained model qualitatively and quantitatively validated that our model can appropriately predict pedestrian behaviors.

The rest of this paper is organized as follows.
Section \ref{sect:model_description} describes the multi scale model of pedestrian flows proposed in this study. 
Section \ref{sect:model_training} presents the details of the training data and learning results.
Section \ref{sect:validation_of_simulation} simulated pedestrian flows at Shibuya Scramble Crossing using the trained model.
In Section \ref{sect:discussion}, the results are discussed, and finally Section \ref{sect:conclusion} concludes the paper.

\begin{figure}[t]
  \centering
  \begin{subfigure}{0.5\linewidth}
    \centering
    \includegraphics[width=\linewidth]{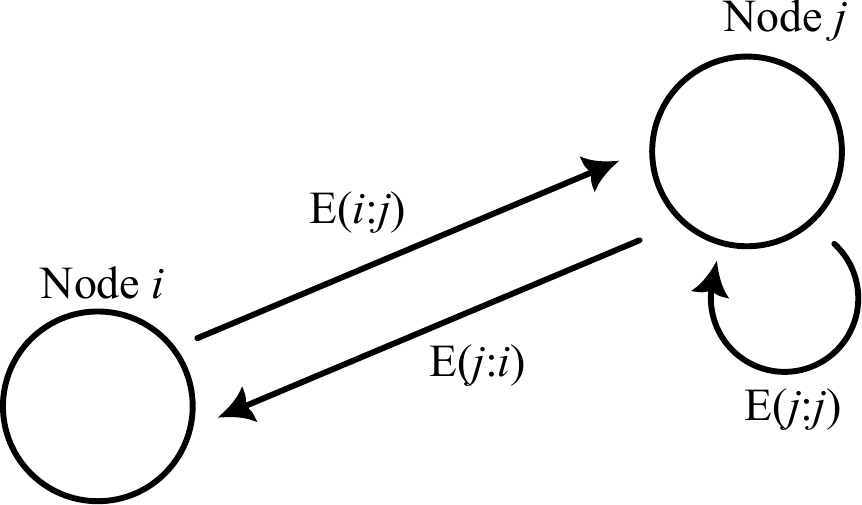}
    \caption{}
    \label{fig:sub:GENERAL_GLOBAL_GRAPH}
  \end{subfigure}
  
  \vspace{2mm}
  
  \begin{subfigure}{0.7\linewidth}
    \centering
    \includegraphics[width=\linewidth]{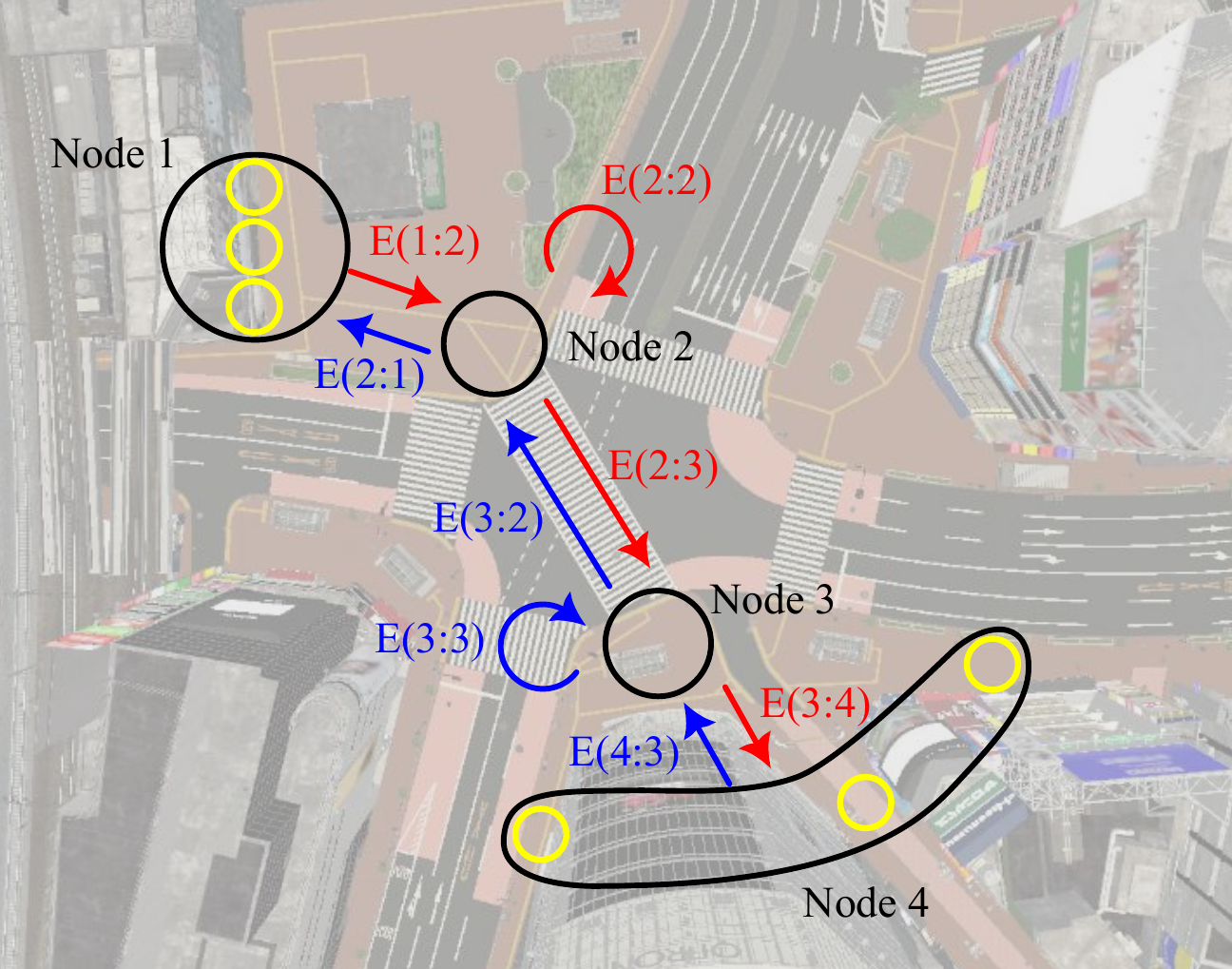}
    \caption{}
    \label{fig:sub:SHIBUYA_GRAPH}
  \end{subfigure}

  \caption{Graph structure representing global-scale behavior of route selection. (a) A destination is represented as a node, and a route between two nodes as an edge. (b) Graph structure setting at Shibuya scramble crossing. The focus is on two pedestrian flows indicated by red and blue edges connecting four nodes.}
  \label{fig:GLOBAL_MOVE_AND_SHIBUYA}
\end{figure}
\begin{figure*}[tbp]
    \centering
    \includegraphics[width=0.8\hsize]{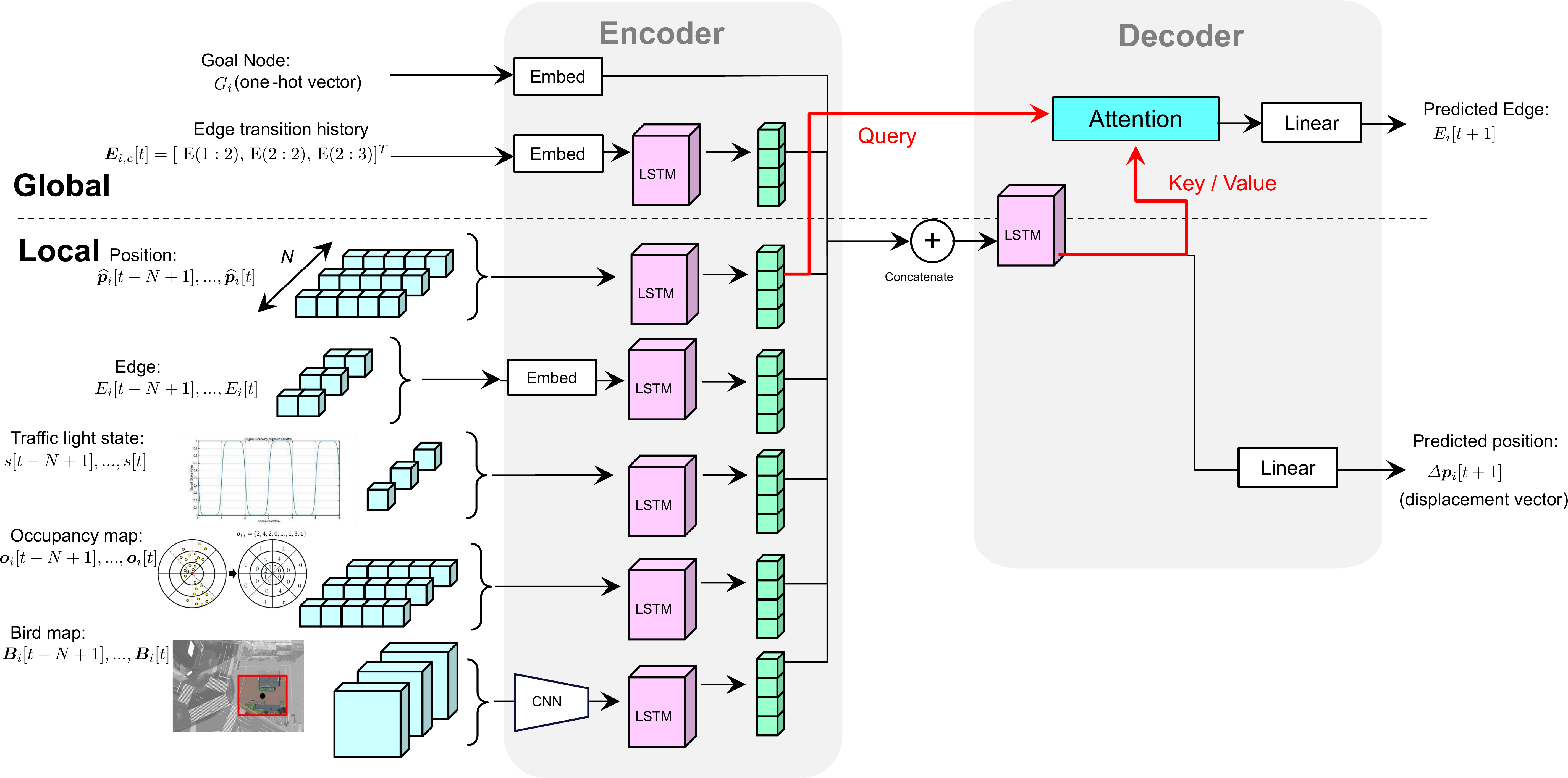}
    \caption{Learning model using LSTM and Attention mechanism for modeling global and local pedestrian behaviors．The encoder/decoder structure uses LSTM, in which spatio-temporal multi scale information is converted into hidden states and integrated
    in the middle layer. The decoder outputs predicted values of pedestrian position, $\varDelta \bm{p}[t+1]$, and edge in the graph structure, $E[t+1]$, in the next time step. Attention mechanism guarantees the consistency between predicted position and edge.}
    \label{fig:LSTM_MODEL_STRUCTURE}
\end{figure*}
\begin{figure*}[tbp]
  \centering
  \begin{minipage}{0.3\linewidth}
    \centering
    \includegraphics[height=28mm]{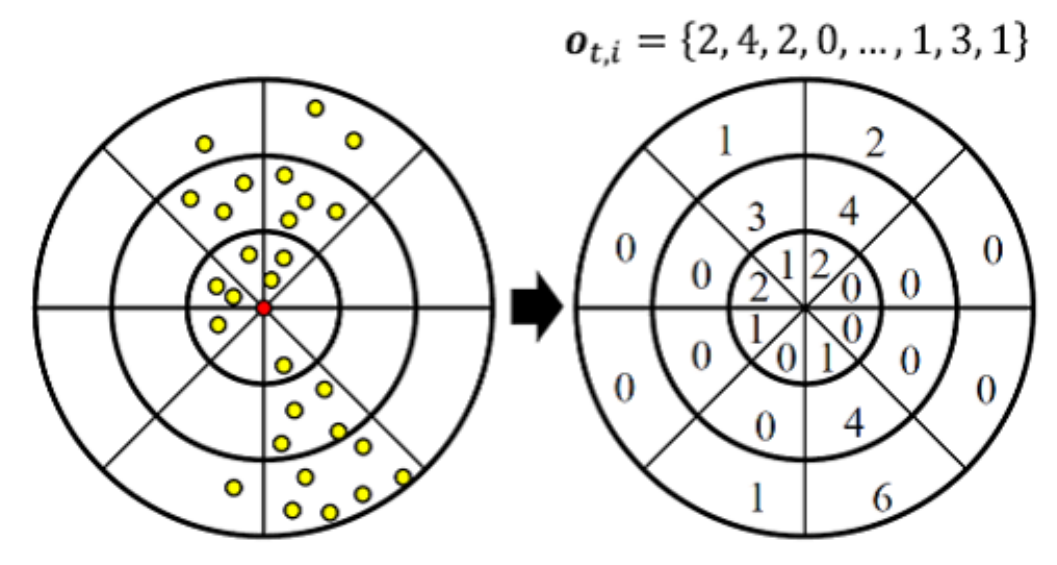}
    \subcaption{Occupancy map}
    \label{fig:sub:OCCUPANCY_MAP_EXAMPLE}
  \end{minipage}
  \hfill
  \begin{minipage}{0.3\linewidth}
    \centering
    \includegraphics[height=28mm]{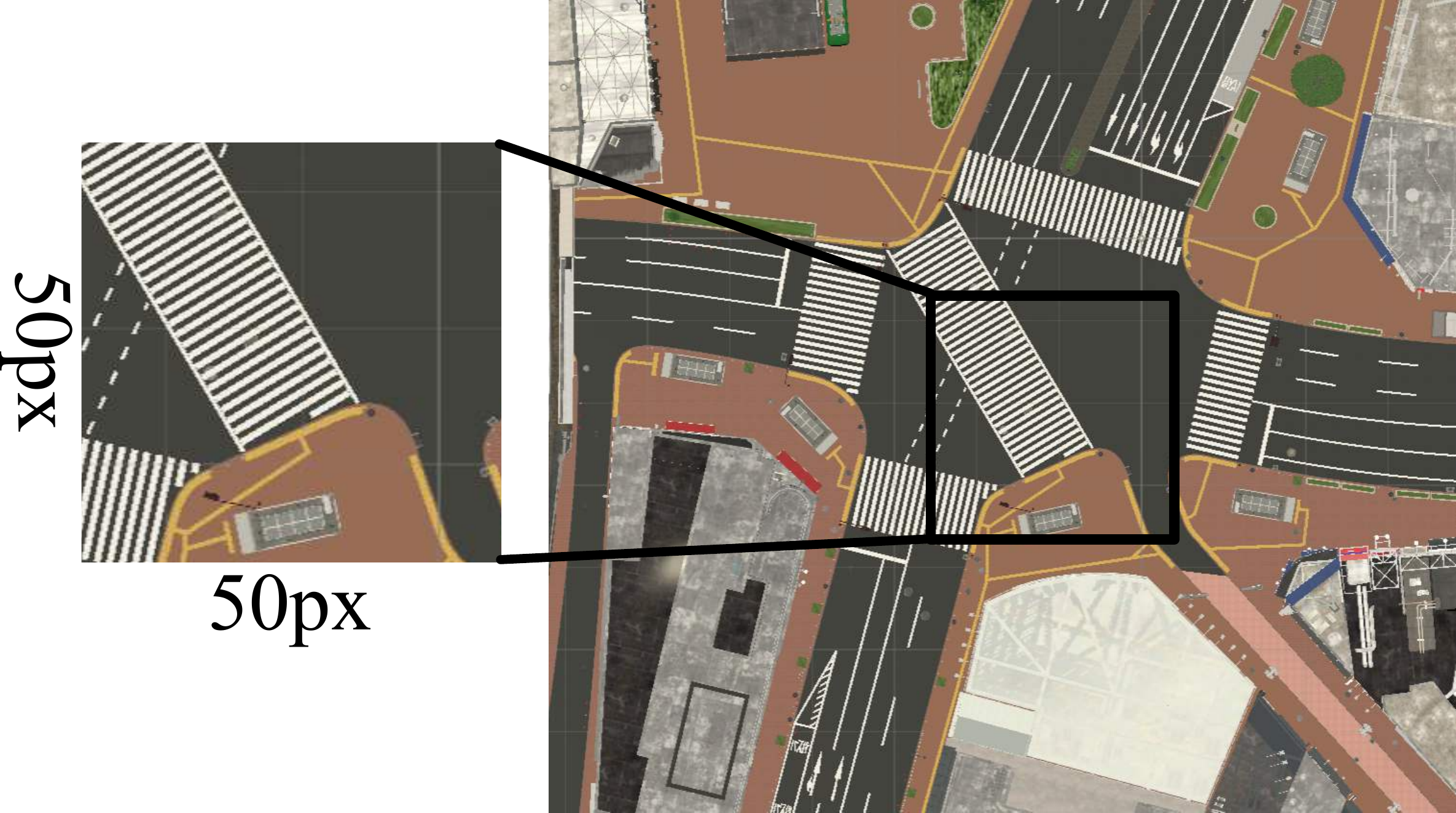}
    \subcaption{Bird map}
    \label{fig:sub:BIRDS_EYE_VIEW_EXAMPLE}
  \end{minipage}
  \hfill
  \begin{minipage}{0.3\linewidth}
    \centering
    \includegraphics[height=28mm]{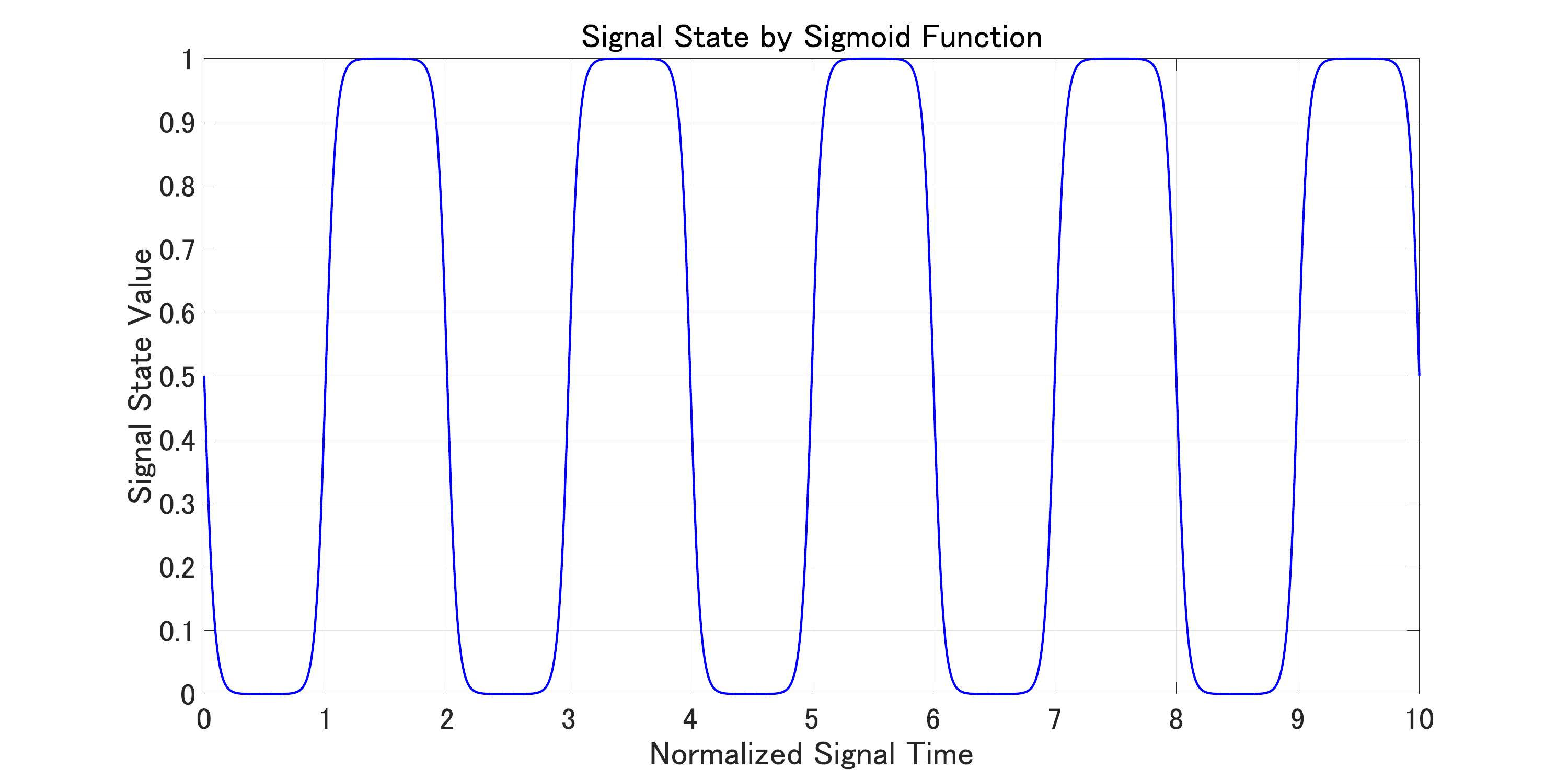}
    \subcaption{Traffic light state represented by a sigmoid function}
    \label{fig:sub:SIGMOID_SIGNAL_SAMPLE}
  \end{minipage}
  \caption{Input information of surrounding environment. 
  (a) Occupancy map \cite{Xue2018} is a circular grid around the target pedestrian (indicated by red). The number of neighbors (indicated by yellow) is counted and input as a constant-size vector $\bm{o}_{t,i}$. 
  (b) Bird map \cite{Xue2018} with 50$\times$50 pixel size is also input as environmental information. 
  (c) Traffic light state is represented by a sigmoid function, in which $s[t] = 0$ and 1 stand for the red and green light states, respectively.}
  \label{fig:MODEL_INPUT_EXAMPLES}
\end{figure*}
\section{Multi scale Model of Pedestrian Flows in Scramble Crossing}
\label{sect:model_description}
\subsection{Problem Setting: Graph Structure of Global Route Selection}
\label{sect:sub:problem_setting}
In this study, we considered a pedestrian movement on the 2D plane and focused on the following global and local behaviors:
\begin{itemize}
    \item{\bf Global behavior:} route selection to a destination, which has a spatial scale of several meters and temporal scale of several tens of seconds.
    \item{\bf Local behaviors:} collision avoidance with other pedestrians and actions based on the surrounding environmental information on a spatial scale of several tens of cm and temporal scale of several seconds.
\end{itemize}
The route selection is represented by the graph structure shown in Fig. \ref{fig:GLOBAL_MOVE_AND_SHIBUYA}\subref{fig:sub:GENERAL_GLOBAL_GRAPH}.
Each location is represented as Node $i$. A directed edge from Node $i$ to $j$ is denoted by E$(i:j)$, and the opposite direction edge is denoted by E$(j:i)$.
We assumed that pedestrians select an edge based on the global and local information and move between nodes.
This representation considers pedestrian behavior such as stopping at a red traffic light by being on {\it loop} edge E$(j:j)$, which means that the pedestrian stays at the same node.

A 3D model of Shibuya scramble crossing (NoneCG, Tokyo Shibuya 3D Model) and its graph structure that we focused on are shown in Fig. \ref{fig:GLOBAL_MOVE_AND_SHIBUYA}\subref{fig:sub:SHIBUYA_GRAPH}.
The graph structure consists of four nodes: Node 1 is located in front of the Shibuya station, Nodes 2 and 3 are located at both ends of the diagonal crosswalk, and Node 4 covers three entrances of shopping streets.
We focused on the two pedestrian flows indicated by the red and blue routes in Fig. \ref{fig:GLOBAL_MOVE_AND_SHIBUYA}\subref{fig:sub:SHIBUYA_GRAPH}, which are denoted as Flows 1 and 2, respectively.
Edge transitions of these flows are defined as follows:
\begin{itemize}
    \item{\bf Flow 1:}
    E$(1:2) \rightarrow$ E$(2:2) \rightarrow$ E$(2:3) \rightarrow$ E$(3:4)$
    \item{\bf Flow 2:}
    E$(4:2) \rightarrow$ E$(3:3) \rightarrow$ E$(3:2) \rightarrow$ E$(2:1)$
\end{itemize}
Note that E(2:2) and E(3:3) are loop edges for the traffic light.

\subsection{LSTM-based Multi scale Model Integrating Local and Global Behaviors}
\label{sect:sub:model_structure}
To integrate global and local behaviors of pedestrians, we propose the learning model shown in Fig. \ref{fig:LSTM_MODEL_STRUCTURE}.
In this model, the encoder and decoder structure uses LSTM, whereby the LSTM encoder converts geometrically global and local information to hidden states.
Combining the information in the intermediate state, this model outputs the prediction considering the global and local information interactions.

We selected the following information as the input for pedestrian $i$ and the surroundings at the current time step $t$.
\subsubsection{Global-scale inputs}
\begin{itemize}
    \item $G_i$: final destination node of pedestrian $i$.
    \item $\bm{E}_{i,c}[t]$: edge transition history until current time $t$. For example, in Fig. \ref{fig:GLOBAL_MOVE_AND_SHIBUYA}\subref{fig:sub:SHIBUYA_GRAPH}, if the edge route of pedestrian $i$ is E$(1:2)$, E$(2:2)$, and E$(2:3)$, this input results in $\bm{E}_{i,c}[t]=[$ E$(1:2)$$,$ E$(2:2)$$,$ E$(2:3)$$]^T$. 
    This is the largest scale variable on both spatial and temporal scales, that does not explicitly include the temporal information.
\end{itemize}

\subsubsection{Local-scale inputs}
\begin{itemize}    
    \item $\bm{\widehat{p}}_i[t]$: relative position vector from each node position.
    \item $\bm{o}_i[t]$: occupancy map around pedestrian $i$, proposed in SS-LSTM\cite{Xue2018}. Fig. \ref{fig:MODEL_INPUT_EXAMPLES}\subref{fig:sub:OCCUPANCY_MAP_EXAMPLE} shows a circular grid map around the target pedestrian (indicated by red). The number of neighbors (indicated by yellow) is counted and input as a constant-size vector $\bm{o}_{i}[t]$.
    \item $\bm{B}_i[t]$: bird map around pedestrian $i$, which was also used in SS-LSTM\cite{Xue2018}. We used the photo information of the surrounding terrain, buildings, and other environmental features, obtained as a 3D tensor, as shown in Fig. \ref{fig:MODEL_INPUT_EXAMPLES}\subref{fig:sub:BIRDS_EYE_VIEW_EXAMPLE}.
    \item $E_i[t]$: an edge that pedestrian $i$ belongs to.
    \item $s[t]$: state value of the traffic light, in which 0 and 1 represent the red and green lights, respectively. To reproduce the passage of time at the crosswalk traffic light, this value is represented using a sigmoid function as shown in Fig. \ref{fig:MODEL_INPUT_EXAMPLES}\subref{fig:sub:SIGMOID_SIGNAL_SAMPLE}.
\end{itemize}
The local-scale inputs included not only the values at the current time-step $t$ but also past $N$ step time-series data: $\ast[t], \cdots, \ast[t-N+1]$.
Compared to the edge history used as the global-scale input, these time-series data have more instantaneous information.
In the next section, we employ 5 FPS data and set $N = 20$, for the input data to include past 4 s of information.
This combination of temporally long and short information also contributes to the integration of global and local behaviors.

\subsubsection{Global- and Local-scale Outputs}
The proposed model outputs the following global- and local-scale information for prediction at the next time step $t+1$.
\begin{itemize}
    \item $E_i[t+1]$: an edge on pedestrian $i$ belongs to
    \item $\varDelta\bm{p}_i[t+1]$: displacement vector of pedestrian $i$ from previous time $t$.
\end{itemize}
Using the obtained $\varDelta\bm{p}_i[t+1]$, the position of pedestrian $i$ can be predicted as follows:
\begin{align}
    \bm{p}_i[t+1]= \bm{p}_i[t] + \varDelta\bm{p}_i[t+1] \label{eq:sim_update}
\end{align}
The pedestrian movement can be simulated by iteratively updating its position using \eqref{eq:sim_update}.

In summary, the proposed model integrates geometrically global and local information using the encoder-decoder structure, and the input data also include temporally long and short information.
Therefore, this model can deal with this kind of spatio-temporally multi-scalability.

\begin{figure*}[t]
    \begin{tabular}{cc}
      \begin{minipage}[t]{0.45\hsize}
        \centering
        \includegraphics[keepaspectratio, height=36mm]{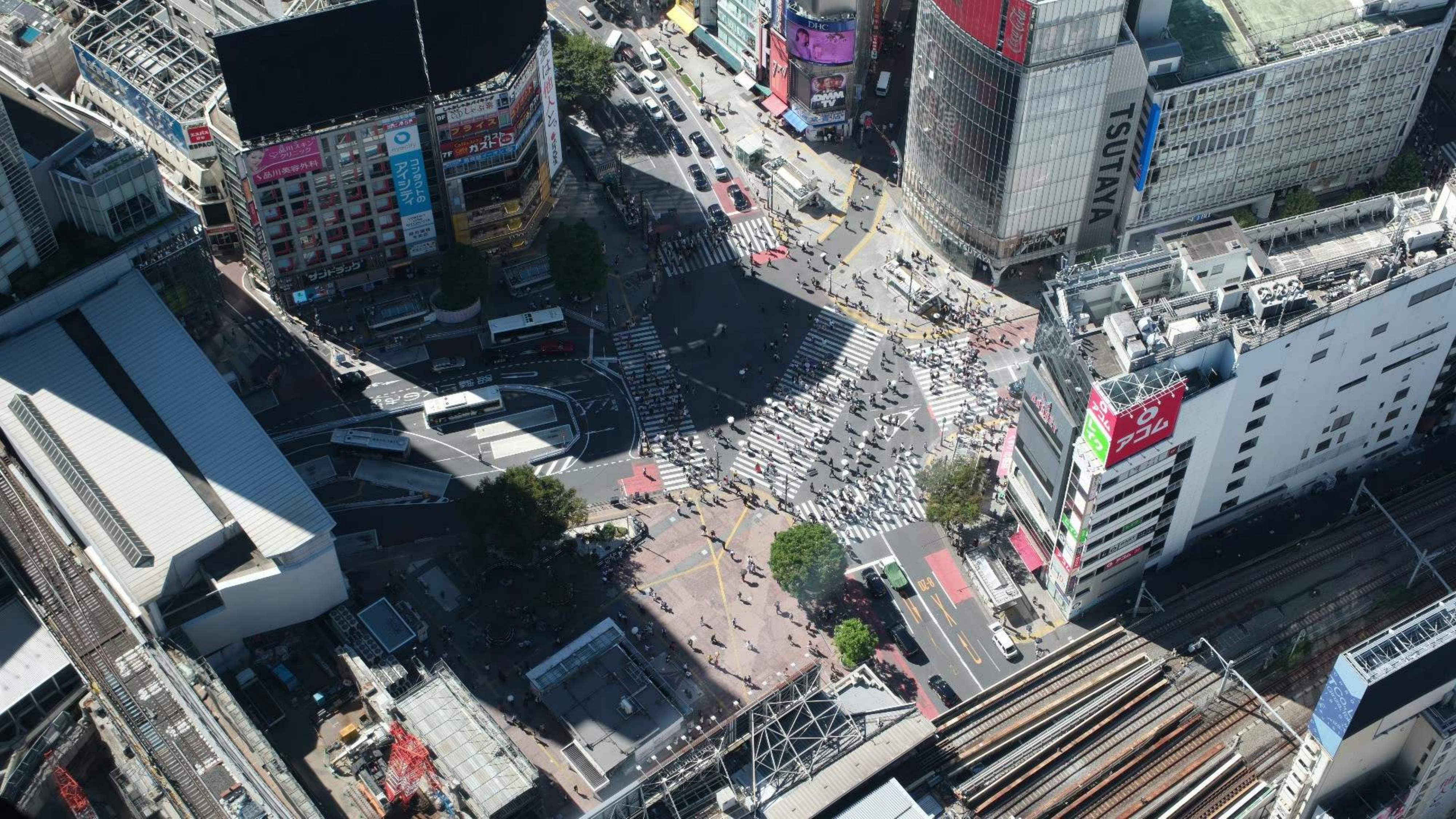}
        \subcaption{}
        \label{fig:SHIBUYA_IMAGE}
      \end{minipage} &
      \begin{minipage}[t]{0.45\hsize}
        \centering
        \includegraphics[keepaspectratio, height=36mm]{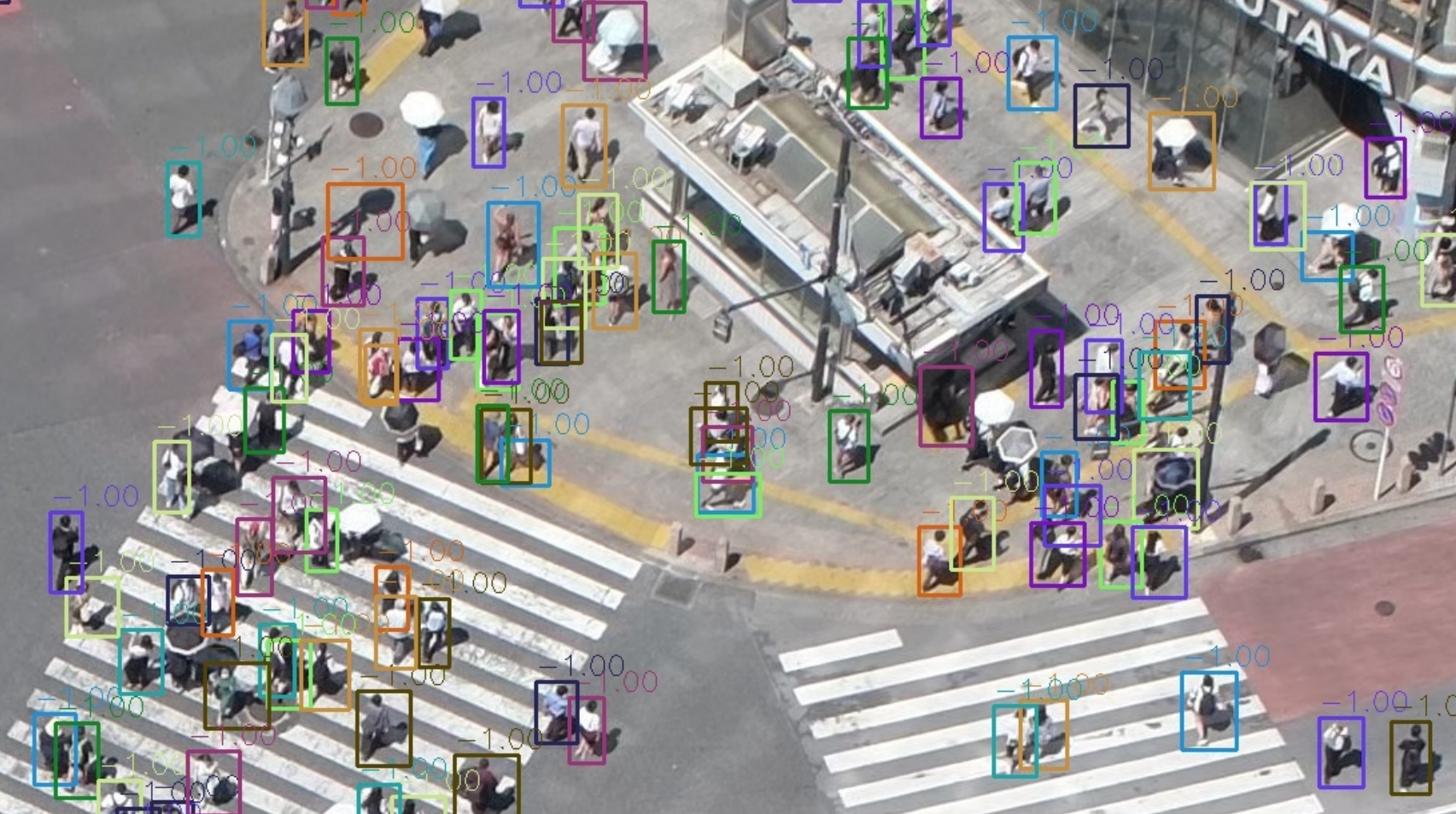}
        \subcaption{}
        \label{fig:SMILETRACK_RESULT}
      \end{minipage} \\
   
      \begin{minipage}[t]{0.45\hsize}
        \centering
        \includegraphics[keepaspectratio, height=36mm]{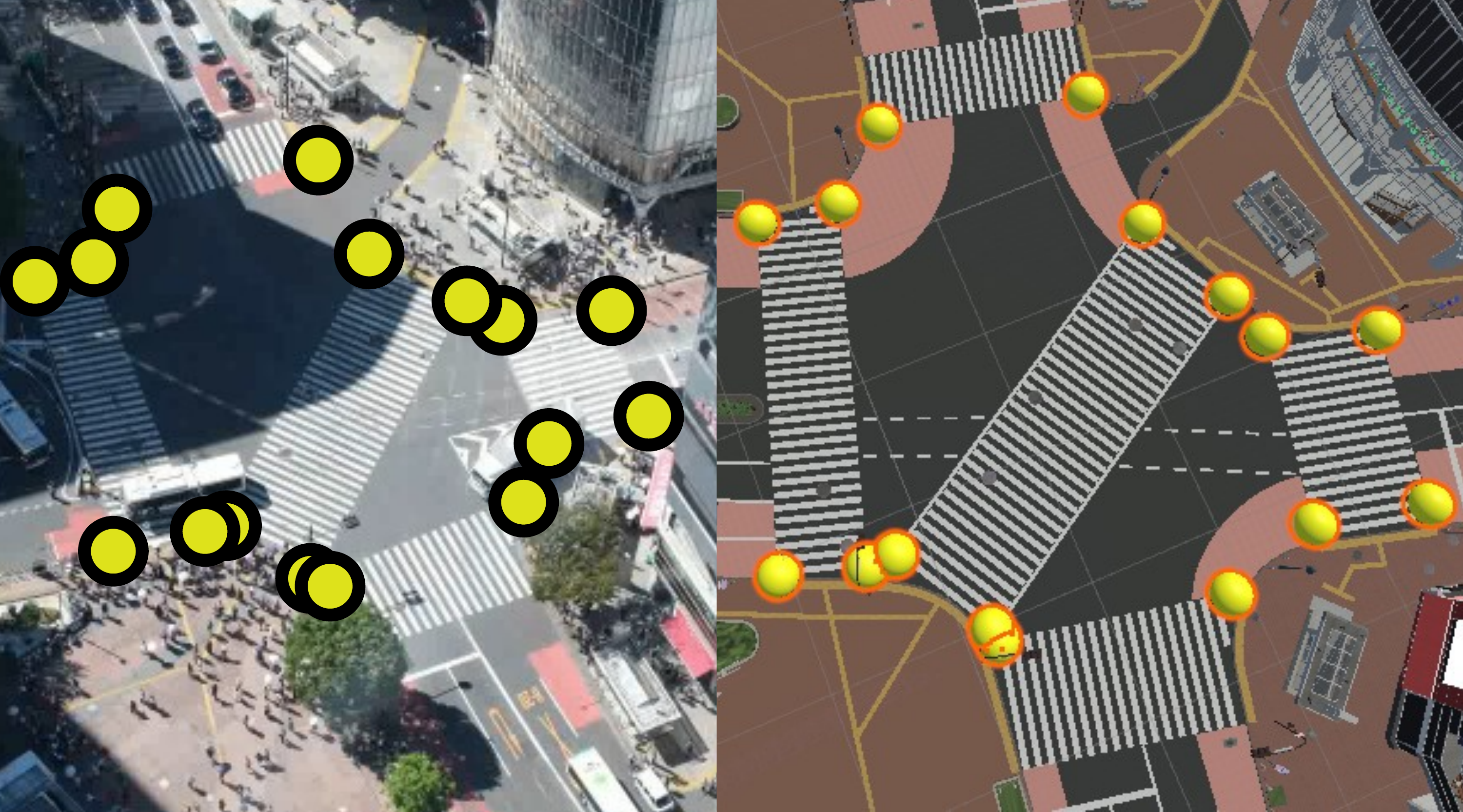}
        \subcaption{}
        \label{fig:ZEBRA_POINTS}
      \end{minipage} &
      \begin{minipage}[t]{0.45\hsize}
        \centering
        \includegraphics[keepaspectratio, height=36mm]{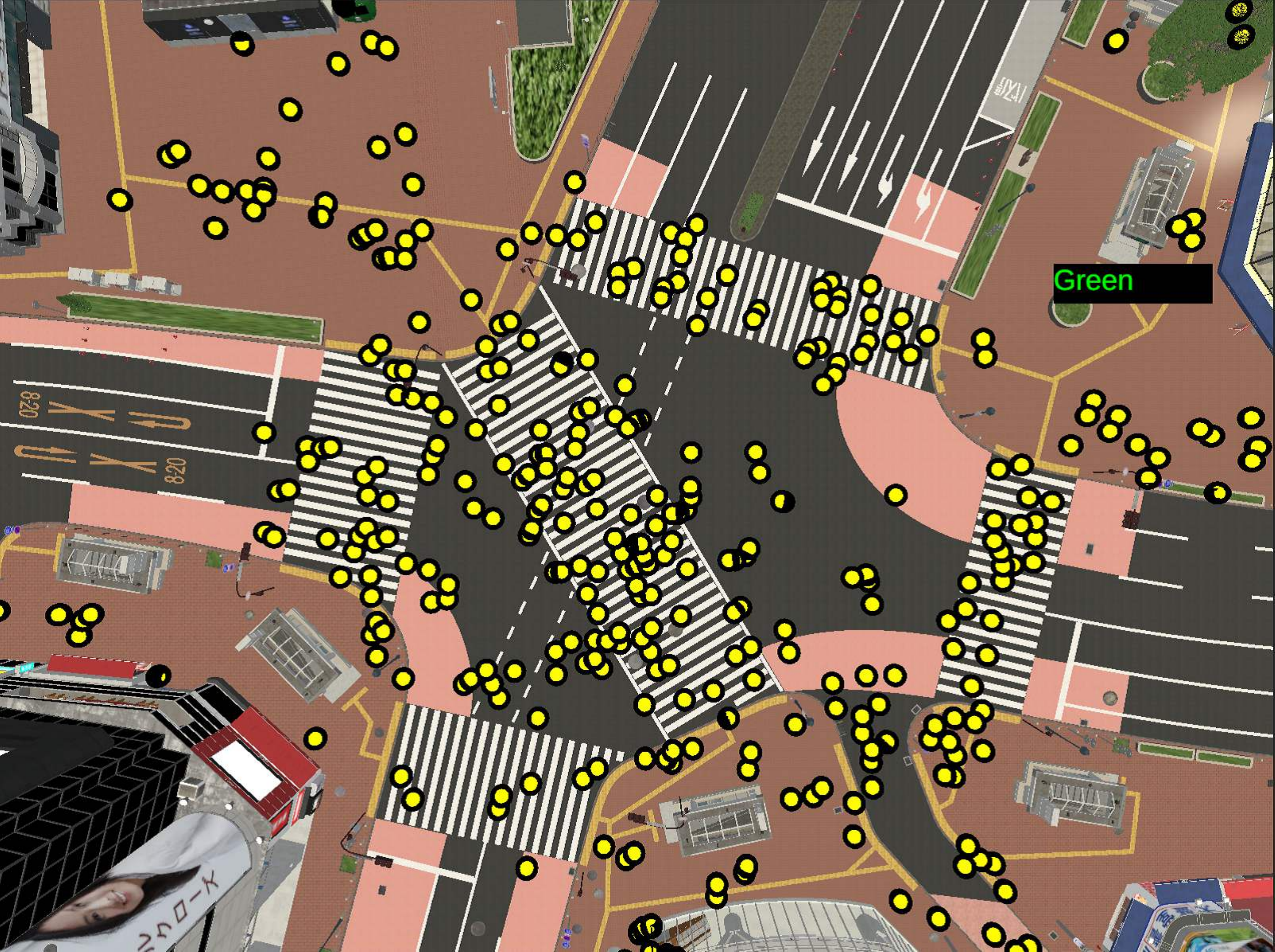}
        \subcaption{}
        \label{fig:SHIBUYA_NORMAL_DATA}
      \end{minipage} 
    \end{tabular}
     \caption{Acquisition of pedestrian data in Shibuya. (a) Pedestrian flows at Shibuya scramble crossing were recorded from a high-rise building using a 8K digital camera. (b) Pedestrians were detected and tracked by SMILETrack \cite{wang2022smiletrack}, as indicated by the bounding boxes. (c) Camera projection matrix $\bm{M}$ was calibrated using selected sample locations. (d) Final 2D position data of pedestrians were obtained using the calibrated matrix.}
  \label{fig:TRANSLATION_PROCESS}
  \end{figure*}
\subsection{Attention Mechanism for Consistency between Global and Local Predictions}
In a preliminary trial \cite{sakurai2024learning}, the proposed model was trained using dummy data generated by the velocity vector field\cite{Yamamoto2013}.
Although the learning achieved good scores (approximately 98 \%\footnote{The accuracy rate of $E_i[t+1]$ was 54.0 \% in \cite{sakurai2024learning} but improved to 98 \% using the negative downsampling presented in Section \ref{sect:sub:ANNOTATION_SAMPLING}.} accuracy rate for $E_i[t+1]$ and 0.1 m error for $\varDelta\bm{p}_i[t+1]$), the walking trajectory simulation based on \eqref{eq:sim_update} often generated unnatural behaviors.
For example, we observed that a simulated pedestrian did not stop at the crosswalk when the traffic light was red or failed to start walking when the traffic light turned green.
We also observed that the simulated pedestrian did not follow the route, e.g., walking straight even when the direction should have been changed from E(1:2) to E(2:3).
This result suggests that simply concatenating global and local hidden states is insufficient to learn the consistency between predicted local positions and global edges.

Therefore, we employed Attention mechanism \cite{vaswani2017attention} to maintain the consistency between the global edge state and local position of the pedestrian.
Attention mechanism is a system that assigns weights to and emphasizes particularly important parts within input data.
Inputs of the mechanism include Query, Key and Value.
Query specifies what to focus on within the input data, whereas Key/Value represent the features and content of the input data.
Weights are assigned based on the similarity between Query and Key/Value.
As shown in Fig. \ref{fig:LSTM_MODEL_STRUCTURE}, an Attention mechanism was introduced, using the hidden state of the position encoded by LSTM as the Query and decoded intermediate state as the Key/Value.



Note that we also tested a similar model structure, using iTransformer \cite{liu2023itransformer} instead of LSTM because it was reported that iTransformer can improve performance when handling multivariate time series data.
Although iTransformer achieved good scores comparable to those of our model (99.5 \%  accuracy rate of $E_i[t+1]$ and within 0.02 m error of $\varDelta\bm{p}[t+1]$), the simulated pedestrians exhibited unnatural behaviors.
iTransformer also uses an Attention mechanism to assign weights to important information between variables.
However, iTransformer is constructed to calculate correlations across all variables and all time series in the input, making it difficult to learn how to assign these weights effectively.

\begin{table}[t]
    \centering
    \caption{Number of pedestrian trajectories obtained by SMILETrack, interpolated using a trained model and estimated by manual counting.}
\resizebox{\columnwidth}{!}{
    \begin{tabular}{l|l|c|c|c}
         & & Flow 1 & Flow 2 & Total  \\\hline
         \multirow{2}{*}{\begin{tabular}{l}
              50-min.\\data
         \end{tabular}} &\begin{tabular}{l}
              (a) Estimated\\by manual count
         \end{tabular}& $\sim 1000$ & $\sim 1000$ & $\sim 2000$ \\ \cline{2-5}
         & \begin{tabular}{l}
              (b) Fully-tracked\\by SMILETrack\end{tabular} & \begin{tabular}{c} 183\\(18\%)\end{tabular} &\begin{tabular}{c} 224\\(22\%)\end{tabular} & \begin{tabular}{c} 407\\(20\%)\end{tabular} \\\hline\hline
         \multirow{4}{*}{\begin{tabular}{l}
              280-sec.\\data extracted\\from 50-min.\\data
         \end{tabular}} & \begin{tabular}{l}
              (c) Estimated\\by manual count\end{tabular} & $\sim 160$ & $\sim 60$ & $\sim 220$ \\\cline{2-5}
              & \begin{tabular}{l}
              (d) Fully-tracked\\by SMILETrack\end{tabular} & \begin{tabular}{c} 6\\($\sim 4$\%)\end{tabular} & \begin{tabular}{c} 13\\($\sim 21$\%)\end{tabular} & \begin{tabular}{c} 19\\($\sim 8$\%)\end{tabular} \\\cline{2-5}
              & \begin{tabular}{l}
              (e) Interpolated\\in Section \ref{sect:sub:weekday_sim}\end{tabular} &\begin{tabular}{c} 83\\($\sim 52$\%)\end{tabular} &\begin{tabular}{c} 26\\($\sim 43$\%)\end{tabular} &\begin{tabular}{c} 109\\($\sim 50$\%)\end{tabular} \\\cline{2-5}
              & \begin{tabular}{l}
              (f) Used for\\simulation\\((d) + (e))\end{tabular} &\begin{tabular}{c} 89\\($\sim 56$\%)\end{tabular} &\begin{tabular}{c} 39\\($\sim 65$\%)\end{tabular} &\begin{tabular}{c} 128\\($\sim 58$\%)\end{tabular} \\\hline
         
    \end{tabular}
}
\label{tab:OBTAINED_TRAJECTORIES}
\end{table}
\section{Learning and Basic Evaluations}
\label{sect:model_training}
\subsection{Data Acquisition of Pedestrians}
\label{sect:sub:data_acquisition}
A challenge in learning the proposed model is how to obtain the pedestrian data.
One of the standard data formats is origin-destination (OD) data, which is usually commercially available.
The OD data is a time-series data format that divides a map into multiple regions ranging from tens of meters to tens of kilometers in size, showing how many pedestrians moved between each divided region over periods ranging from minutes to hours.
OD data represent the movement of pedestrians between a starting and ending location. 
However, from the perspective of personal information protection, the data cannot be used without statistical processing, and individual routes cannot be tracked.
Additionally, the spatial and temporal resolutions of OD data are approximately 10 m and 10 min, respectively, which are insufficient for training the local behaviors in this study.

Therefore, in this study, we captured pedestrian movements from a building around Shibuya scramble crossing 
using a video camera with 8K resolution and 30 FPS (FUJIFILM, XF16-80mmF4 R OIS WR) for recording and obtained pedestrian position data through image processing.
Fig. \ref{fig:TRANSLATION_PROCESS}\subref{fig:SHIBUYA_IMAGE} shows a recorded camera video image.

The pedestrians in the recorded videos were detected and tracked using SMILETrack \cite{wang2022smiletrack}, as shown in Fig. \ref{fig:TRANSLATION_PROCESS}\subref{fig:SMILETRACK_RESULT}.
SMILETrack outputs ID, bounding box coordinates, and detection confidence scores of the detected pedestrian.
Then, the relationship between the camera images and 3D model is calibrated to obtain the 2D positions in the Shibuya 3D model space shown in Fig. \ref{fig:TRANSLATION_PROCESS}\subref{fig:SHIBUYA_NORMAL_DATA} from tracked pedestrian pixel coordinates in the images.
Let $(u, v)$ denote the pixel coordinate in the camera image, and $(X,Y,Z)$ denote the 3D coordinates in the 3D Shibuya model.
Using the camera projection matrix $\bm{M}$, the following equation holds:
\begin{align}
    s\bm{u} = \bm{M}\bm{x}\label{eq:picTo3D}
\end{align}
where $\bm{u}=[u\,v\,1]^T,\bm{x}=[X\,Y\,Z\,1]^T$, and $s$ denotes a scale parameter.
In this paper, we applied direct linear transformation (DLT) method \cite{Hartley:2003:MVG:861369}, with $\bm{M}$ representing both the camera's orientation/position and its internal parameters in a single linear model.
Using multiple sets of $\bm{u}$ and $\bm{x}$ for corresponding locations in the camera image and 3D model, we can estimate matrix $\bm{M} \in \mathbb{R}^{3 \times 4}$.
The 16 selected locations shown in Fig. \ref{fig:TRANSLATION_PROCESS}\subref{fig:ZEBRA_POINTS}, are vertices of the marked crosswalk lines.
As detailed in Appendix \ref{append:pos3d_calculaion}, 
by determining the value of $Z$, we can convert $(u,v)$ to $(X,Y)$.
Finally, 2D walking trajectories are obtained as pedestrian data of Flows 1 and 2.

\begin{figure}[t]
  \centering
  \begin{subfigure}{0.7\linewidth}
    \centering
    \includegraphics[width=\linewidth]{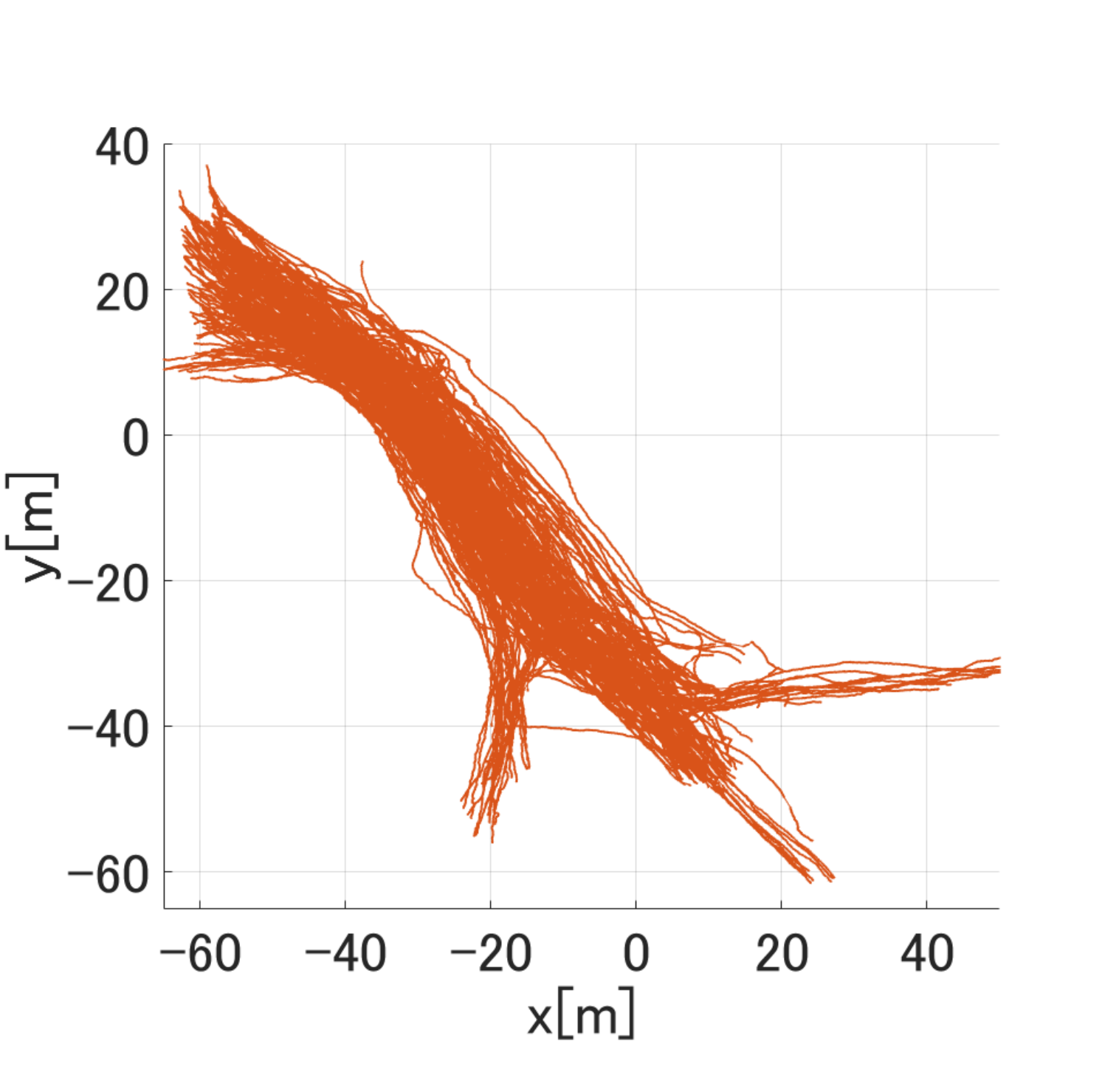}
    \subcaption{Flow 1}
    \label{fig:sub:FLOW1_TRAIN_TRAJ}
  \end{subfigure}
  
  
  \begin{subfigure}{0.7\linewidth}
    \centering
    \includegraphics[width=\linewidth]{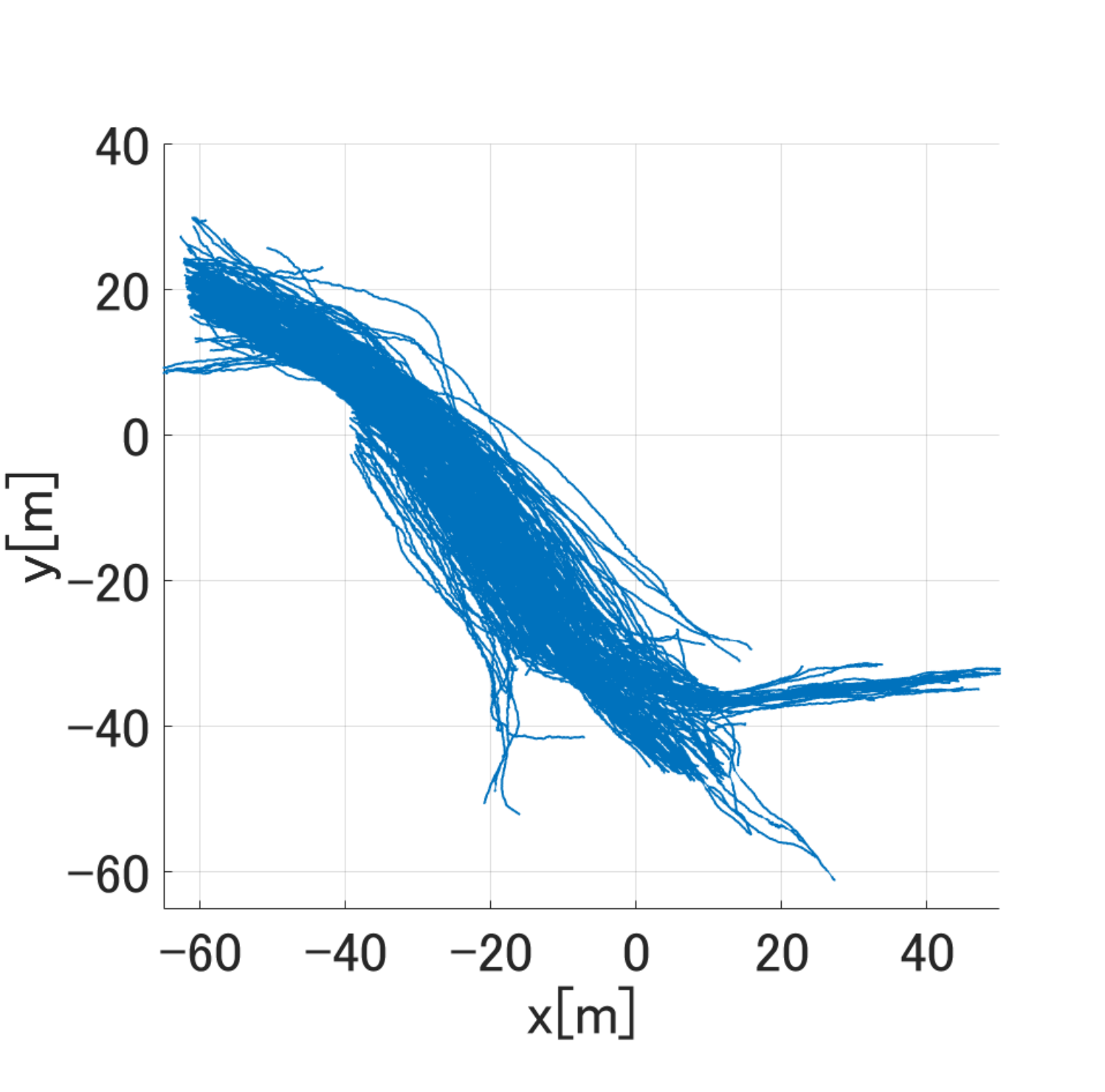}
    \subcaption{Flow 2}
    \label{fig:sub:FLOW2_TRAIN_TRAJ}
  \end{subfigure}

  \caption{All trajectories of (a) Flow 1 and (b) Flow 2 used for training.}
  \label{fig:TRAJ_TRAIN_SOURCE}
\end{figure}

In total, 50-min daytime video data were collected on a weekday, obtaining 407 pedestrians' trajectory data (183 for Flow 1 and 224 for Flow 2), as shown in Fig. \ref{fig:TRAJ_TRAIN_SOURCE}.
The average frame length per 1 pedestrian trajectory was 2591 for Flow 1 and 2260 for Flow 2.

Note that SMILETrack often failed in detecting and tracking pedestrians. 
From the traffic volume in the crosswalk, we estimated the actual number of each pedestrian's flow with 50 minutes approximately corresponding to 1000. 
This means that SMILETrack successfully tracked approximately 18\% of pedestrians for Flow 1 and 22\% for Flow 2.
As discussed in Section \ref{sect:sub:weekday_sim} and Appendix \ref{append:interpolation}, post-processing for trajectory interpolation would be one option, to increase the amount of data for efficient learning. 
In the next section, the raw data without any post-processing is presented for training, aiming to appropriately learn the actual pedestrian behaviors.

\subsection{Edge Annotation and Down Sampling}
\label{sect:sub:ANNOTATION_SAMPLING}
The input and output dataset presented in Section \ref{sect:sub:model_structure} was prepared by manually annotating the obtained trajectories with edge transition for the graph structure.
However, the resultant data became {\it unbalanced} because in the data less time has elapsed at the moment when the value of $E_i[t]$ changes such that $E_i[t] \neq E_i[t+1]$, and  $E_i[t] = E_i[t+1]$ holds for almost all time steps.
A model trained on such a dataset tends to predict the same value with the input value. i.e., $E_i[t+1] = E_i[t]$, making it difficult to predict a sudden change of $E_i[t]$, which is necessary for the edge transition of the route selection.
Therefore, we divided the time-series data into the two groups: one satisfying $E_i[t] = E_i[t+1]$ and the other satisfying $E_i[t] \neq E_i[t+1]$, and then applied negative downsampling \cite{he2009learning} so that each group would have an equal number of data samples.


The number of input-output datasets was 453912 for Flow 1 and 475417 for Flow 2.
Negative downsampling reduced it to 97522 and 104576, respectively, which corresponds to approximately 5-hour data because the frame rate was 30 FPS.

\begin{figure}[t]
 \centering
  \includegraphics[width=0.9\hsize]{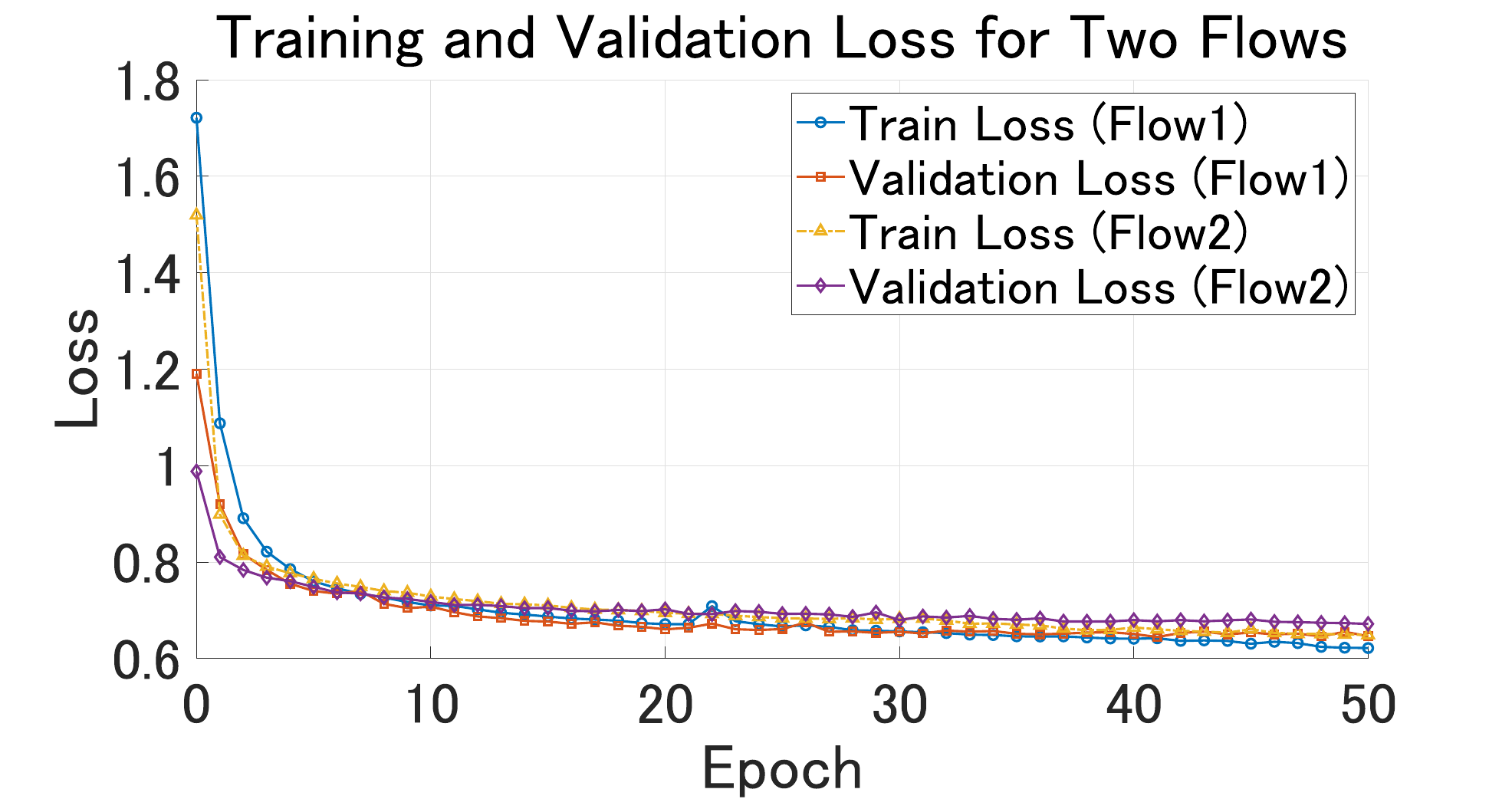}
  \caption{Training and validation loss curves for each flow.}
  \label{fig:TRAINING_LOSS}
\end{figure}
\begin{table}[t]
    \centering
    \caption{Position errors and edge classification accuracies of the models for each flow on the validation dataset.}
    \begin{tabular}{l| c c}
        \toprule
        Flow & Position error [m] & Edge accuracy rate  [\%] \\
        \midrule
        Flow 1 & 0.068 & 99.7 \\
        Flow 2 & 0.070 & 99.5 \\
        \bottomrule
    \end{tabular}
    \label{tab:TRAINING_RESULT}
\end{table}
\subsection{Learning of Multi-scale Model}
The training data were downsampled at 5 FPS during training, validations, and simulations in a manner similar to previous studies \cite{Alahi2016,Xue2018}.
One NVIDIA RTX 3090 GPU was used for training.
The training model was implemented using PyTorch, with a learning rate of $10^{-4}$, using the Adam optimizer and a ReduceLROnPlateau scheduler.
It took approximately 8 hours to complete 50 epochs of training for each flow.
Fig. \ref{fig:TRAINING_LOSS} shows the loss function of each flow.
The horizontal axis represents the number of epochs, and the vertical axis represents the loss.
The blue and yellow lines show the loss on the training data for Flows 1 and 2.
The red and purple lines show the loss on the validation data for Flows 1 and 2, respectively.
Each loss decreased sharply within the first 10 epochs and then continued to decrease gradually until 50 epochs.
Table \ref{tab:TRAINING_RESULT} shows a prediction accuracy of each flow using the model with minimum validation loss.
The position error was less than 0.1 m, and edge prediction achieved an accuracy of more than 99\%.

\begin{figure*}[tbp]
  \centering
  \begin{minipage}{0.24\linewidth}
    \centering
    \includegraphics[width=\linewidth]{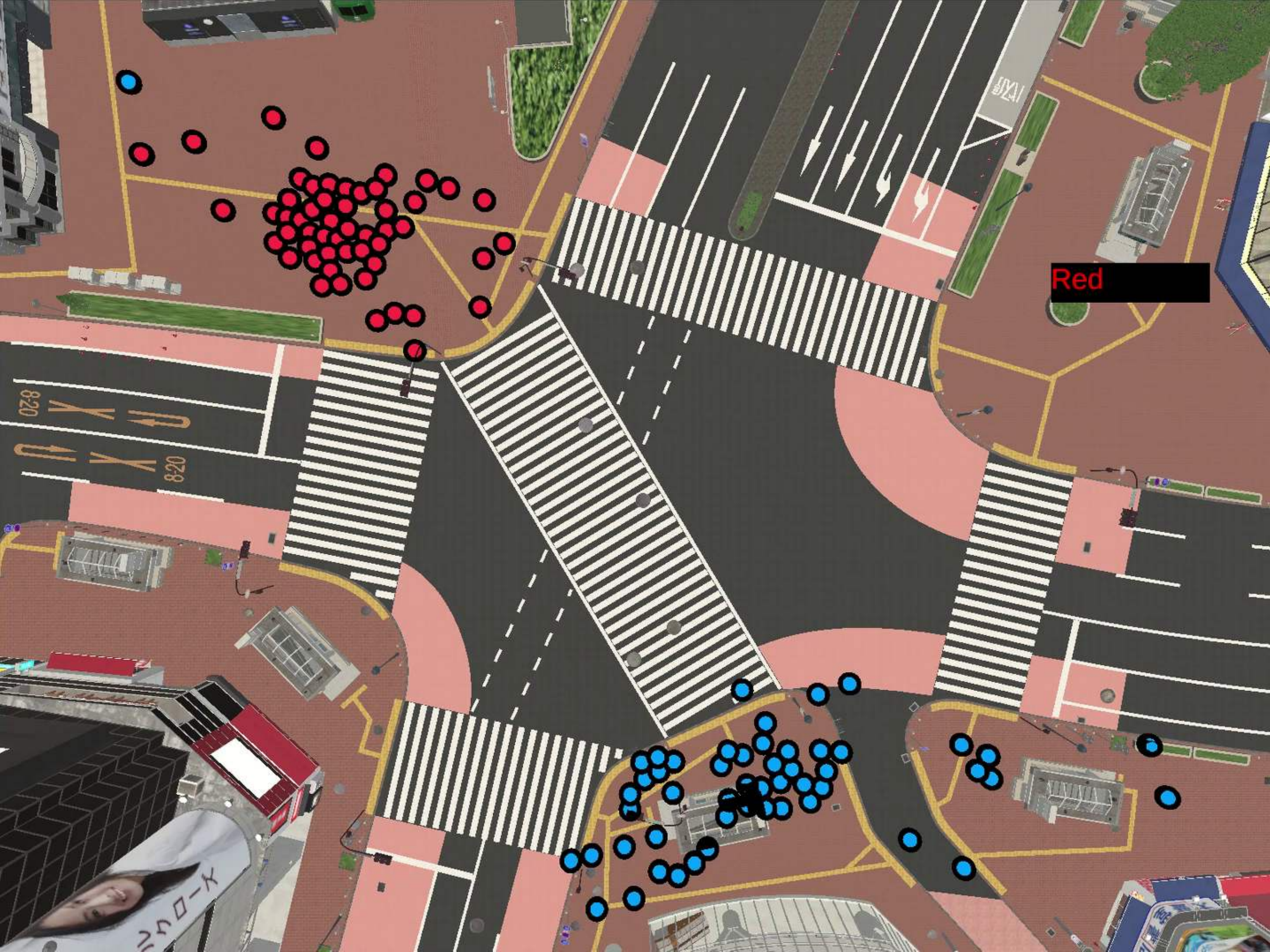}
    \subcaption{}
    \label{fig:sub:COUNTERSIM_RED}
  \end{minipage}
  \hfill
  \begin{minipage}{0.24\linewidth}
    \centering
    \includegraphics[width=\linewidth]{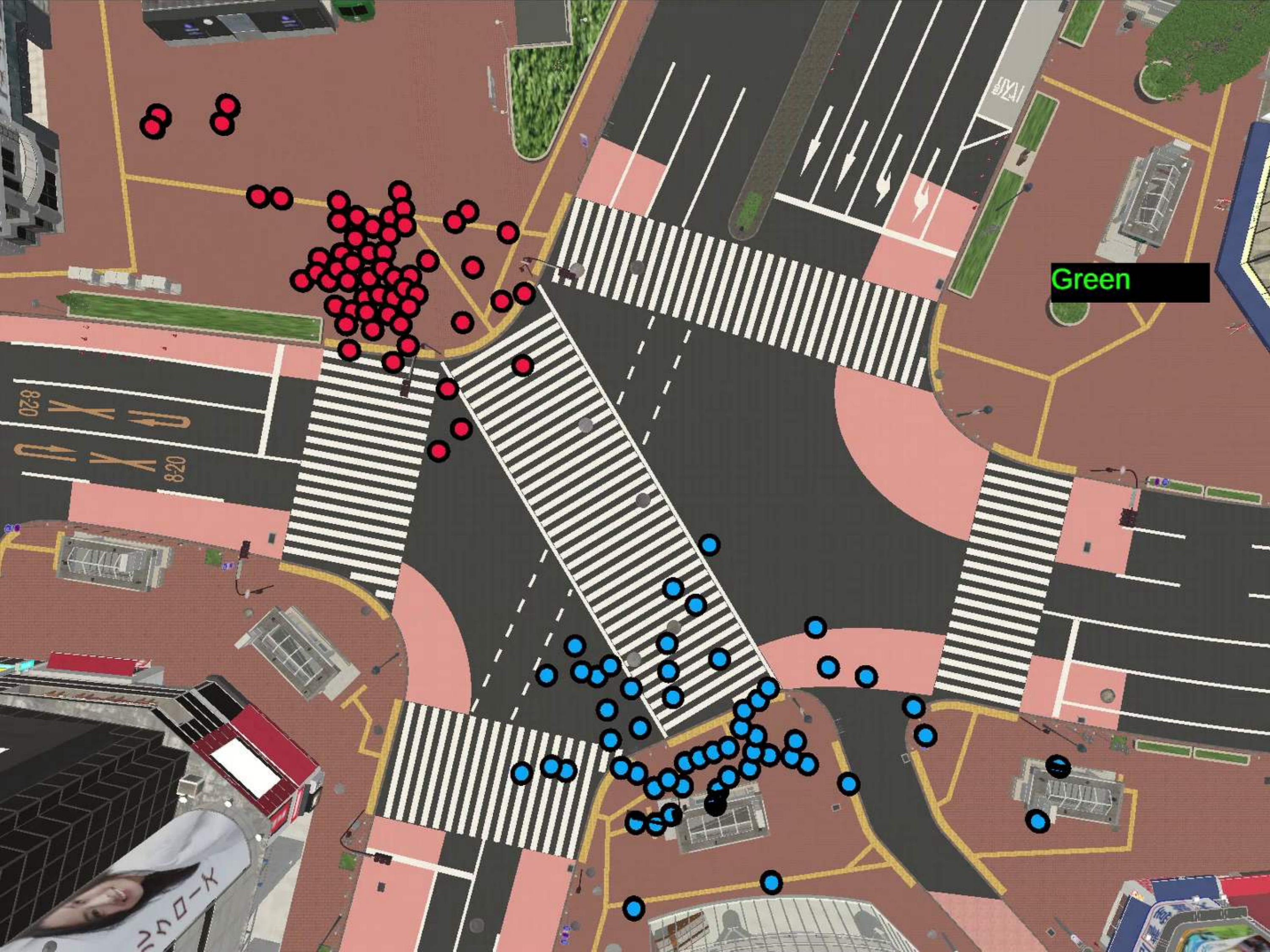}
    \subcaption{}
    \label{fig:sub:COUNTERSIM_ENTER}
  \end{minipage}
  \hfill
  \begin{minipage}{0.24\linewidth}
    \centering
    \includegraphics[width=\linewidth]{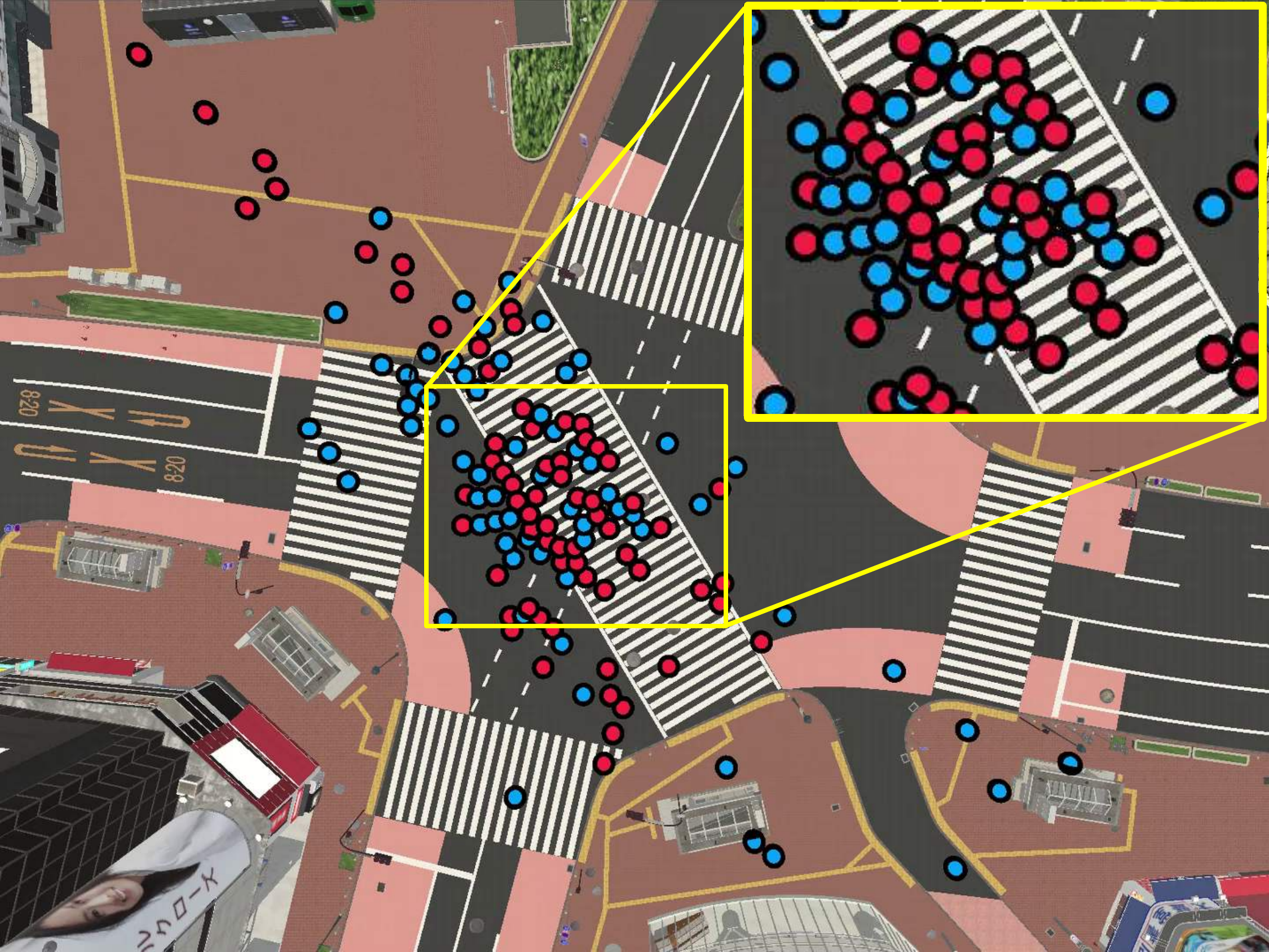}
    \subcaption{}
    \label{fig:sub:COUNTERSIM_LANEFORM}
  \end{minipage}
  \hfill
  \begin{minipage}{0.24\linewidth}
    \centering
    \includegraphics[width=\linewidth]{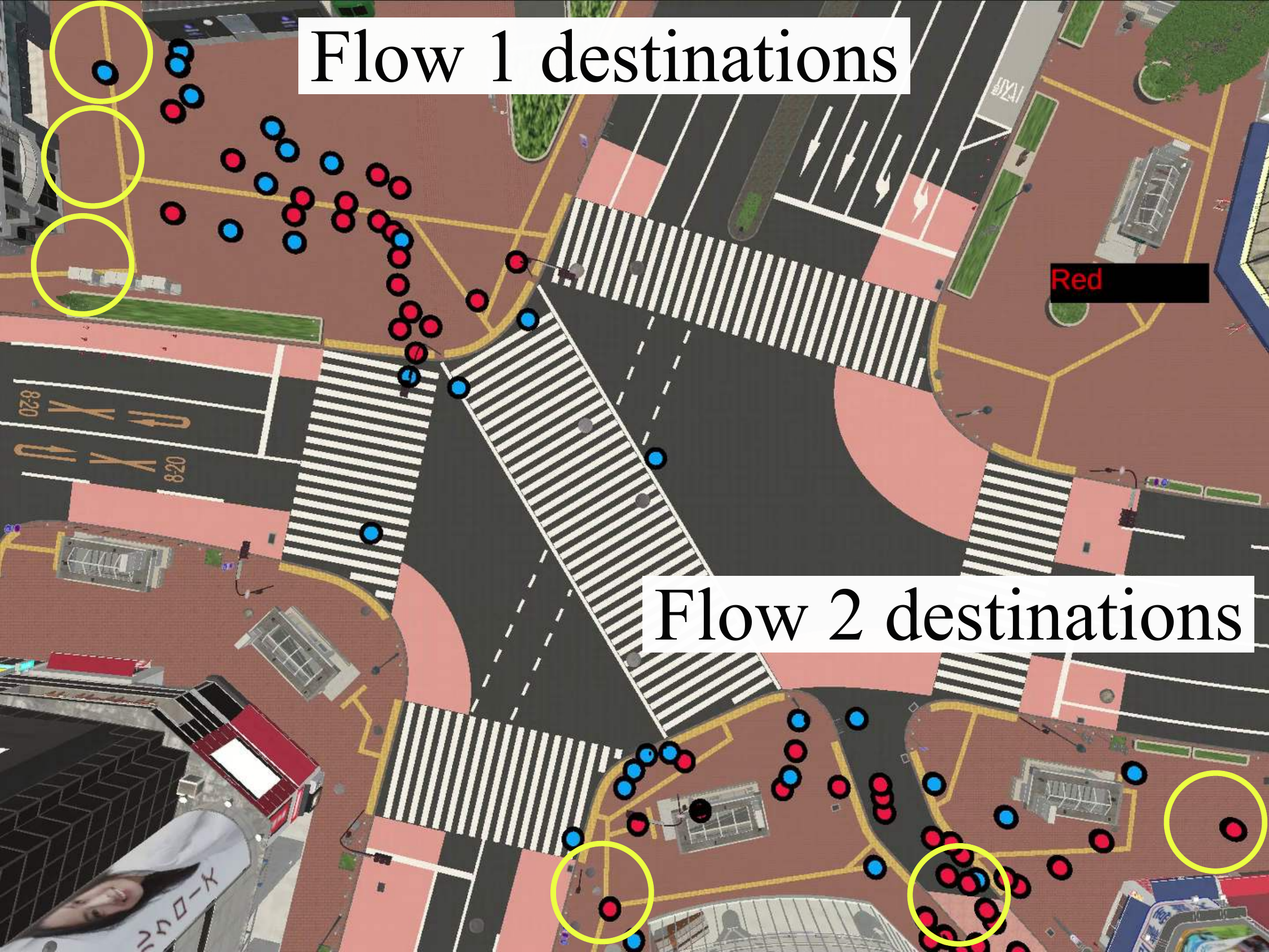}
    \subcaption{}
    \label{fig:sub:COUNTERSIM_DESTINATION}
  \end{minipage}
  \caption{Simulation of Flows 1 (red) and 2 (green) using the trained model: (a) waiting for the traffic light while red, (b) entering the crosswalk, (c) crossing at the crosswalk and (d) reaching the destinations. (c) lane formed by two flows countering as observed in reality \cite{Murakami, helbing1998self}.}
  \label{fig:COUNTERFLOW_SIM}
\end{figure*}
\begin{figure}[tbp]
  \centering
  \begin{minipage}{0.3\linewidth}
    \centering
    \includegraphics[width=\linewidth]{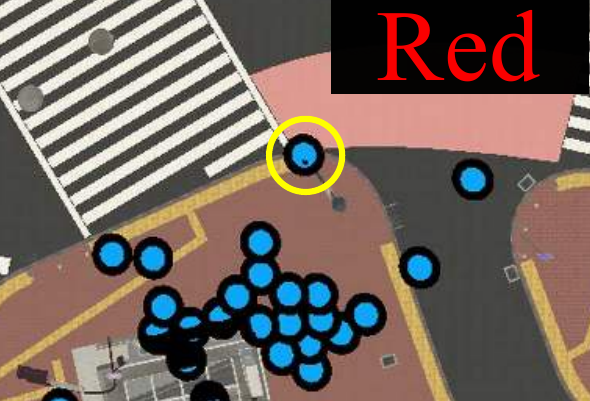}
    \subcaption{11 s}
  \end{minipage}
  \hfill
  \begin{minipage}{0.3\linewidth}
    \centering
    \includegraphics[width=\linewidth]{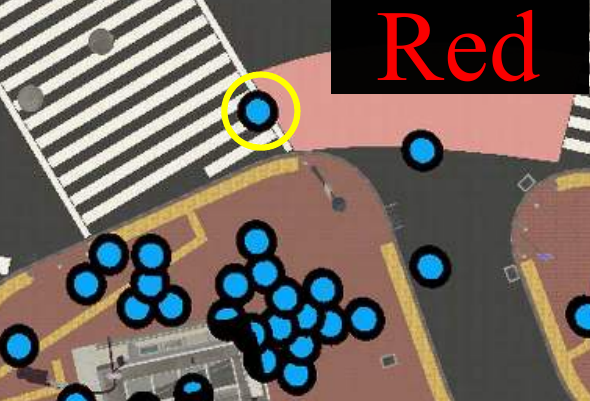}
    \subcaption{12 s}
  \end{minipage}
  \hfill
  \begin{minipage}{0.3\linewidth}
    \centering
    \includegraphics[width=\linewidth]{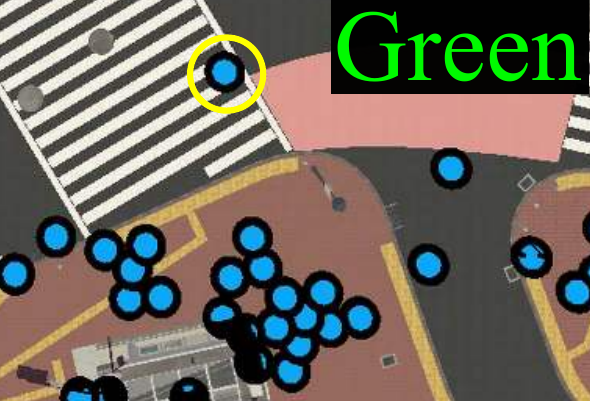}
    \subcaption{13 s}
  \end{minipage}
  \caption{An example of a pedestrian who started to enter the crosswalk just before the traffic light turned green.
  Since such behavior can be observed in the actual data, this result suggests that the proposed model appropriately learned and simulated the pedestrian behaviors in reality.}
  \label{fig:UNDER_RED_SIGNAL_ENTERING}
\end{figure}
\section{Validation of Pedestrian Flow Simulation}
\label{sect:validation_of_simulation}
\subsection{Qualitative Validation}
\label{sect:sub:counterflow_sim}
\subsubsection{Simulation Setting}
We simulated trajectories of virtual pedestrian agents using the trained models in the previous section, and validated that the trained model can reproduce global and local behaviors of the two pedestrian flows.
The procedure of simulating pedestrians in each flow is summarized as follows:
\begin{enumerate}[STEP 1]
    \item Set an initial position of a pedestrian. Randomly select an area from multiple candidates (indicated by the yellow circles in Fig. \ref{fig:GLOBAL_MOVE_AND_SHIBUYA}\subref{fig:sub:SHIBUYA_GRAPH}), and randomly set the initial position within the area. 
    \item Calculate the global and local inputs depending on the pedestrian position and input the information into the trained model.
    \item Update the pedestrian position based on \eqref{eq:sim_update} using the output from the trained model.
    Note that a rule-based collision avoidance effect $\varDelta\bm{p}_{i,a}[t]$ was also added, as explained later.
    Eventually, we used the following updating equation. 
    \begin{align}
        \bm{p}_i[t+1]= \bm{p}_i[t] + \varDelta\bm{p}_i[t+1] + \varDelta\bm{p}_{i,a}[t]\label{eq:sim_update_avoid}
    \end{align}
    \item Go back to STEP 2 and iterate the update of the pedestrian position.
\end{enumerate}

In the initial position setting, for each flow, three candidates locations were set, as shown in Fig. \ref{fig:GLOBAL_MOVE_AND_SHIBUYA}(b).
Flow 1 has three adjoining yellow circle areas at 
Node 1, assuming that pedestrians exit from the entrance of the Shibuya station.
Flow 2 also has three candidate areas at Node 4, but these areas are separately located at the entrance of three different streets. 
Each candidate area has 3 m radius, and
for each flow, a new pedestrian was added every 2 s, to replicate the number of pedestrians at the crosswalk as closely as possible.

When using \eqref{eq:sim_update} to update pedestrian position, we observed that simulated pedestrians sometimes collided with each other because they were approaching more closely than observed in the training data.  
This is because tracking sometimes failed in detecting close neighbors due to occlusion; therefore, the proposed model obtained a collision avoidance behavior for a larger distance but not for a smaller distance. 
In simulations assuming a dense situation, however, sometimes pedestrians approached each other closely as a result of colliding with each other.
In \eqref{eq:sim_update_avoid}, we added $\varDelta\bm{p}_{i,a}[t]$ as an additional collision avoidance effect based on the following equations \cite{Yamamoto2013}:
\begin{align}
    \varDelta\bm{p}_{i,a}[t] &= \sum_{i\neq j} s\left(\|\bm{r}_{ij}\|\right)\frac{\bm{r}_{ij}}{\|\bm{r}_{ij}\|} \\
    \bm{r}_{ij} & := \bm{p}_j[t] - \bm{p}_i[t] \\
    s(r) & := \frac{c}{1 + \exp(a(r - b))}
\end{align}
where $s(r)$ is a sigmoid function, and the parameter $b$ represents a personal-space distance, which is the distance within which collision avoidance becomes effective.
In this study, we set these parameters as $a=10$, $b=0.8$, and $c=2.5$ by trial and error.

\subsubsection{Result}
Fig. \ref{fig:COUNTERFLOW_SIM} shows snapshots of the simulation result for 280 s.
In the figure,  red and blue points indicate the position of the simulated pedestrians for Flows 1 and 2, respectively.
The text in the upper right of each snapshot indicates the traffic light status.
We implemented the simulations using Unity considering future virtual reality (VR) simulator applications \cite{sakurai2023vr}.
The Unity side performed pedestrian rendering and input data acquisition, and the Python side predicted the pedestrian output using a learning model constructed with PyTorch.
Input/Output data transmission was conducted via shared memory.

In Fig. \ref{fig:COUNTERFLOW_SIM}\subref{fig:sub:COUNTERSIM_RED}, the pedestrians of both flows are waiting in front of the crosswalk in accordance with the red traffic light.
In Fig. \ref{fig:COUNTERFLOW_SIM}\subref{fig:sub:COUNTERSIM_ENTER}, the pedestrians start to walk when the traffic light turns green.
In particular, in Fig. \ref{fig:UNDER_RED_SIGNAL_ENTERING} and the attached video, some pedestrians begin crossing just before the traffic light turns green.
Since such behavior can be observed in actual data, this result suggests that the proposed model appropriately learned and simulated real-life pedestrian behaviors, representing the traffic light using the sigmoid function, as shown in Fig. \ref{fig:MODEL_INPUT_EXAMPLES}\subref{fig:sub:SIGMOID_SIGNAL_SAMPLE}.

In Fig. \ref{fig:COUNTERFLOW_SIM}\subref{fig:sub:COUNTERSIM_LANEFORM}, pedestrians of Flows 1 and 2 are crossing each other in the middle of the crosswalk, and 
a lane forms \cite{Murakami, helbing1998self}, which is a self-organizing phenomenon seen in counter flows.
Finally, in Fig. \ref{fig:COUNTERFLOW_SIM}\subref{fig:sub:COUNTERSIM_DESTINATION},pedestrians of each flow reach their destination after crossing. 
From these results, we can qualitatively validate that the trained model appropriately reproduced the global route selection, including the behaviors according to traffic light. 

In particular, it is interesting that pedestrians of Flow 1 part ways and head off on three different paths without colliding with buildings although no explicit information was provided for such branch trajectories, only the topological information of Node 4.
This result suggests that combining topological route information with graph structure and local environmental data allows the proposed model to appropriately obtain such realistic behavior.


\begin{figure}[t]
  \centering
  \begin{subfigure}{0.7\linewidth}
    \centering
    \includegraphics[width=\linewidth]{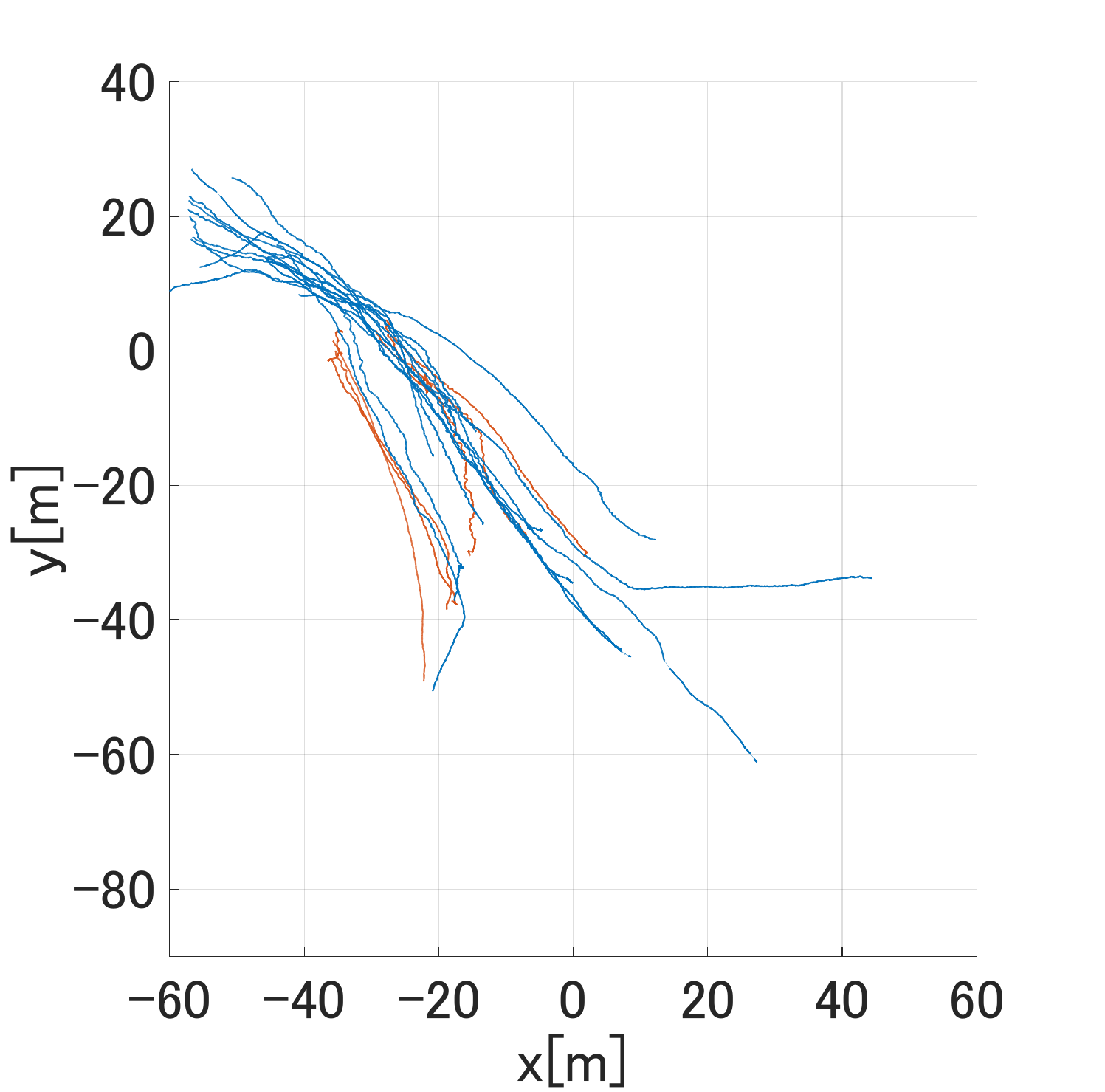}
    \subcaption{Trajectories used for training}
    \label{fig:sub:TRAJ_TRAIN_SOURCE}
  \end{subfigure}
  
  \vspace{2mm}
  
  \begin{subfigure}{0.7\linewidth}
    \centering
    \includegraphics[width=\linewidth]{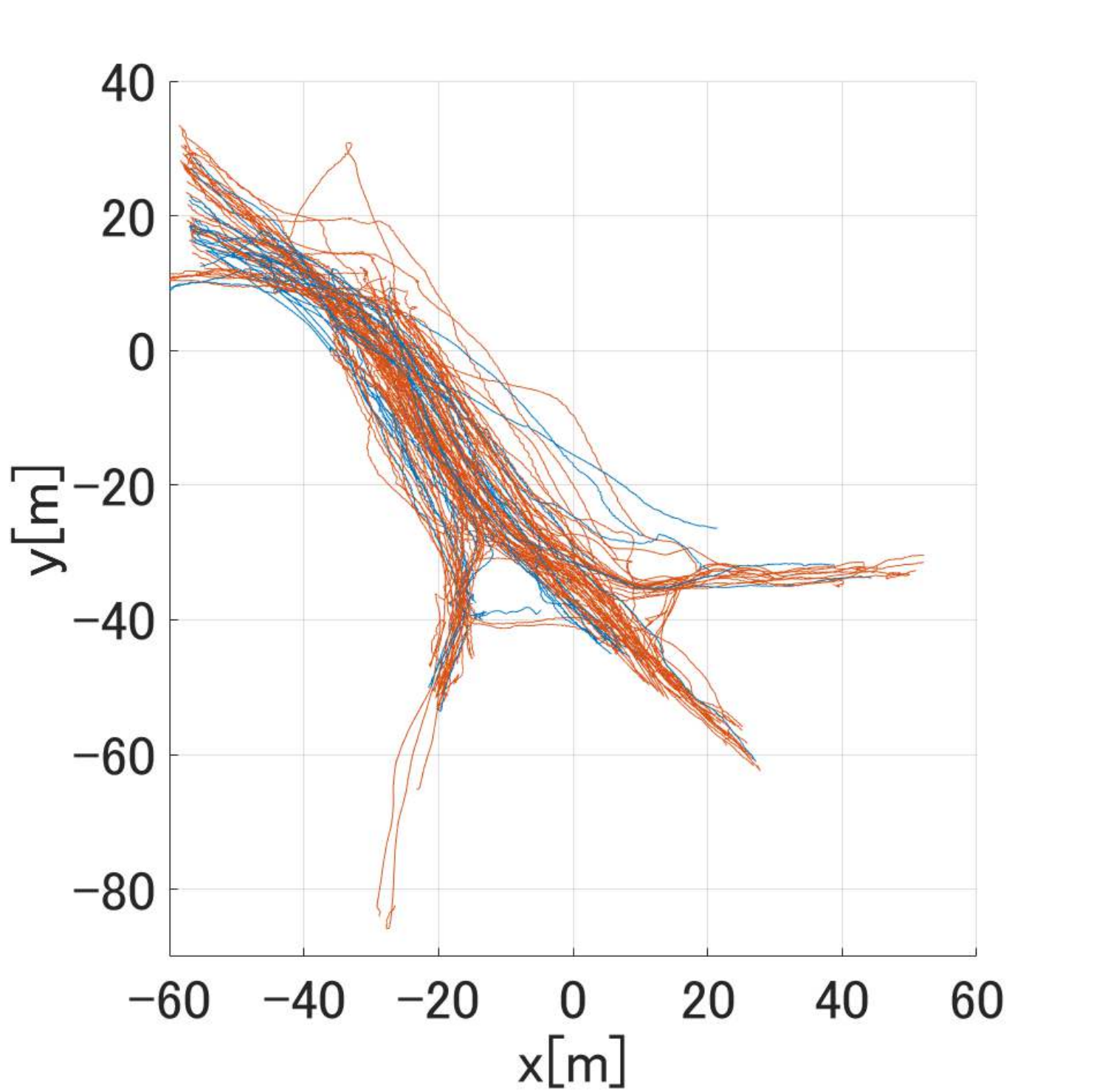}
    \subcaption{Interpolated trajectories}
    \label{fig:sub:INTERPOLATED_TRAJECTORIES}    
  \end{subfigure}

  \caption{Trajectories used for the Section \ref{sect:sub:weekday_sim} simulation. Red trajectories denote Flow 1 and blue trajectories denote Flow 2. (a) Trajectories of the two flows were used for training using the simulation period extracted from Figs. \ref{fig:TRAJ_TRAIN_SOURCE}\subref{fig:sub:FLOW1_TRAIN_TRAJ} and \subref{fig:sub:FLOW2_TRAIN_TRAJ}, obtaining only 19 pedestrian trajectories. (b) 109 consistent trajectories obtained through prediction-based interpolation on the weekday data.
  }
  \label{fig:WEEKDAY_REPLACED_TRAJECTORIES}
\end{figure}
\begin{figure*}[t]
    \begin{tabular}{cc}
      \begin{minipage}[t]{0.45\hsize}
        \centering
        \includegraphics[keepaspectratio, width=0.9\linewidth]{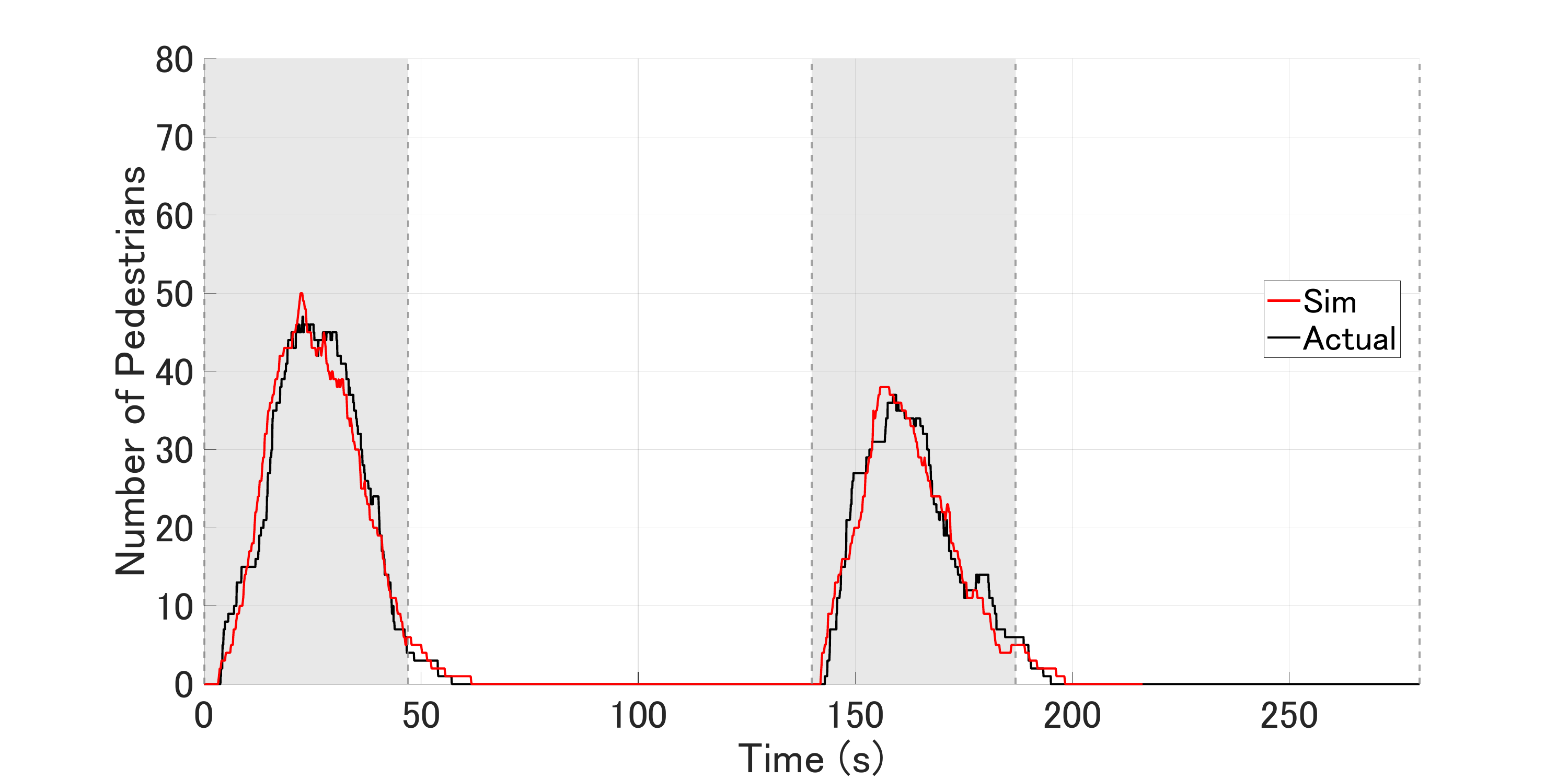}
        \subcaption{Number of Flow 1}
        \label{fig:sub:WEEKDAY_NUMBER_FLOW1}
      \end{minipage} &
      \begin{minipage}[t]{0.45\hsize}
        \centering
        \includegraphics[keepaspectratio, width=0.9\linewidth]{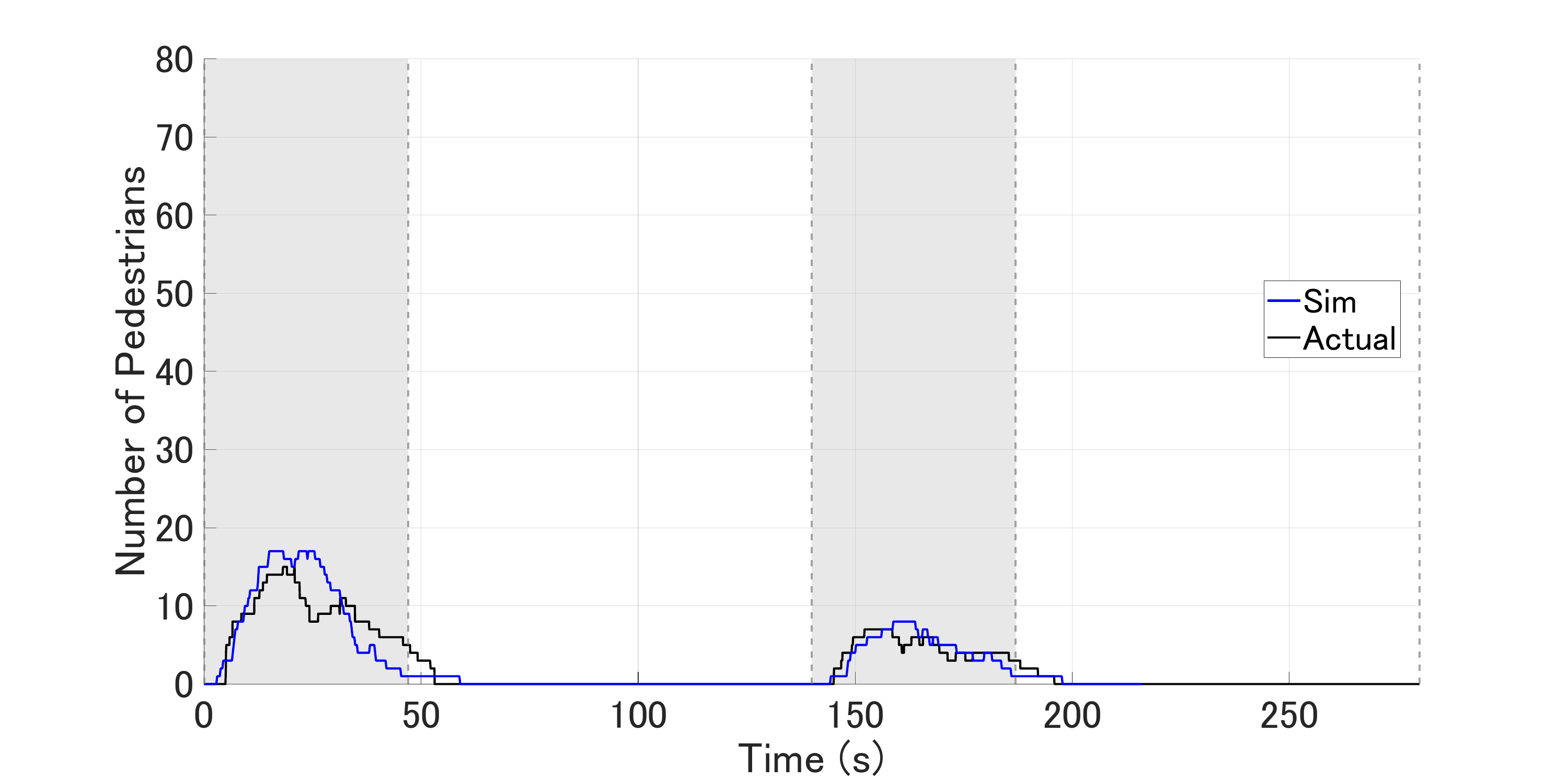}
        \subcaption{Number of Flow 2}
        \label{fig:sub:WEEKDAY_NUMBER_FLOW2}
      \end{minipage} \\
   
      \begin{minipage}[t]{0.45\hsize}
        \centering
        \includegraphics[keepaspectratio, width=0.9\linewidth]{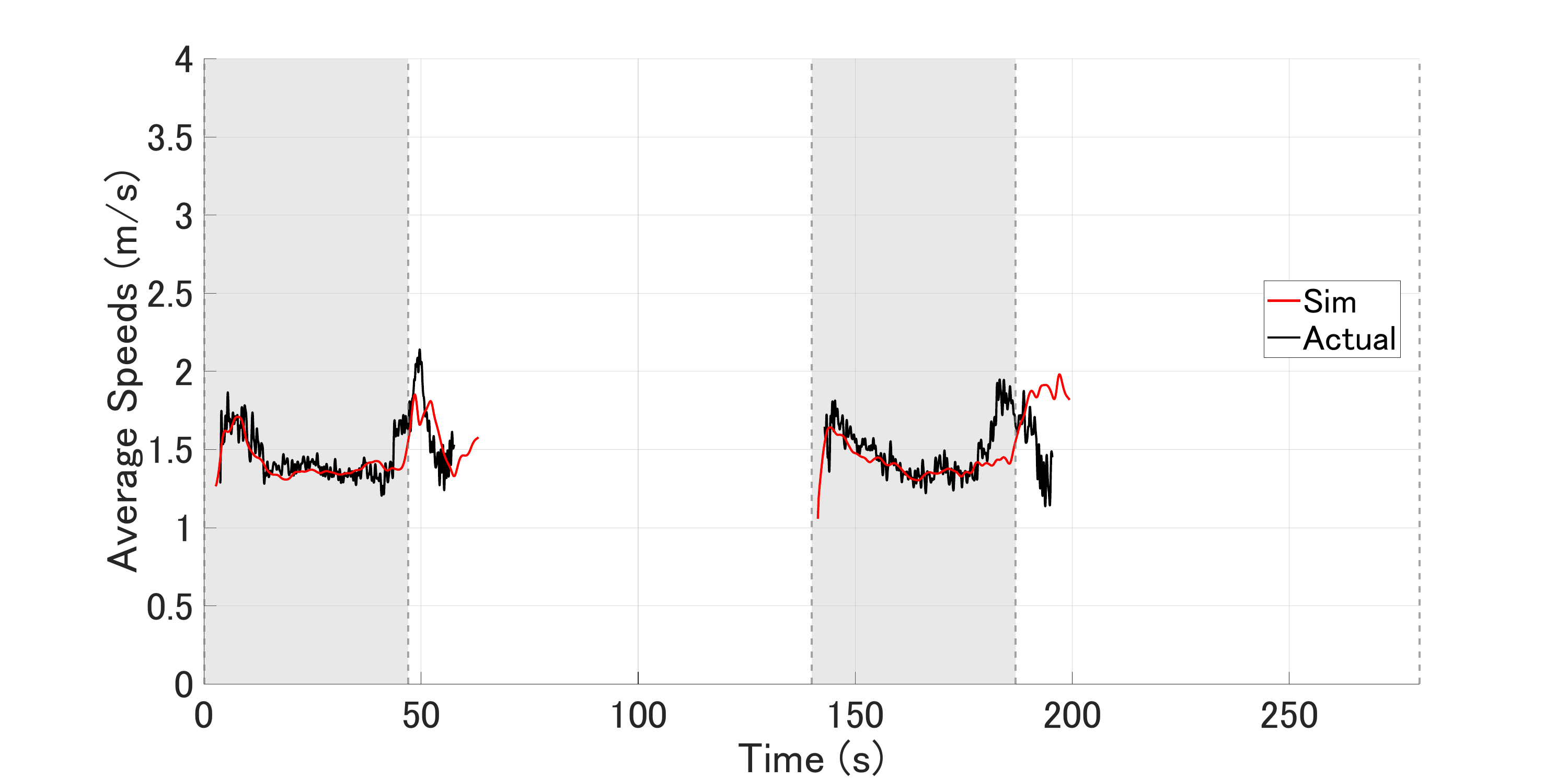}
        \subcaption{Speed of Flow 1}
        \label{fig:sub:WEEKDAY_SPEED_FLOW1}
      \end{minipage} &
      \begin{minipage}[t]{0.45\hsize}
        \centering
        \includegraphics[keepaspectratio, width=0.9\linewidth]{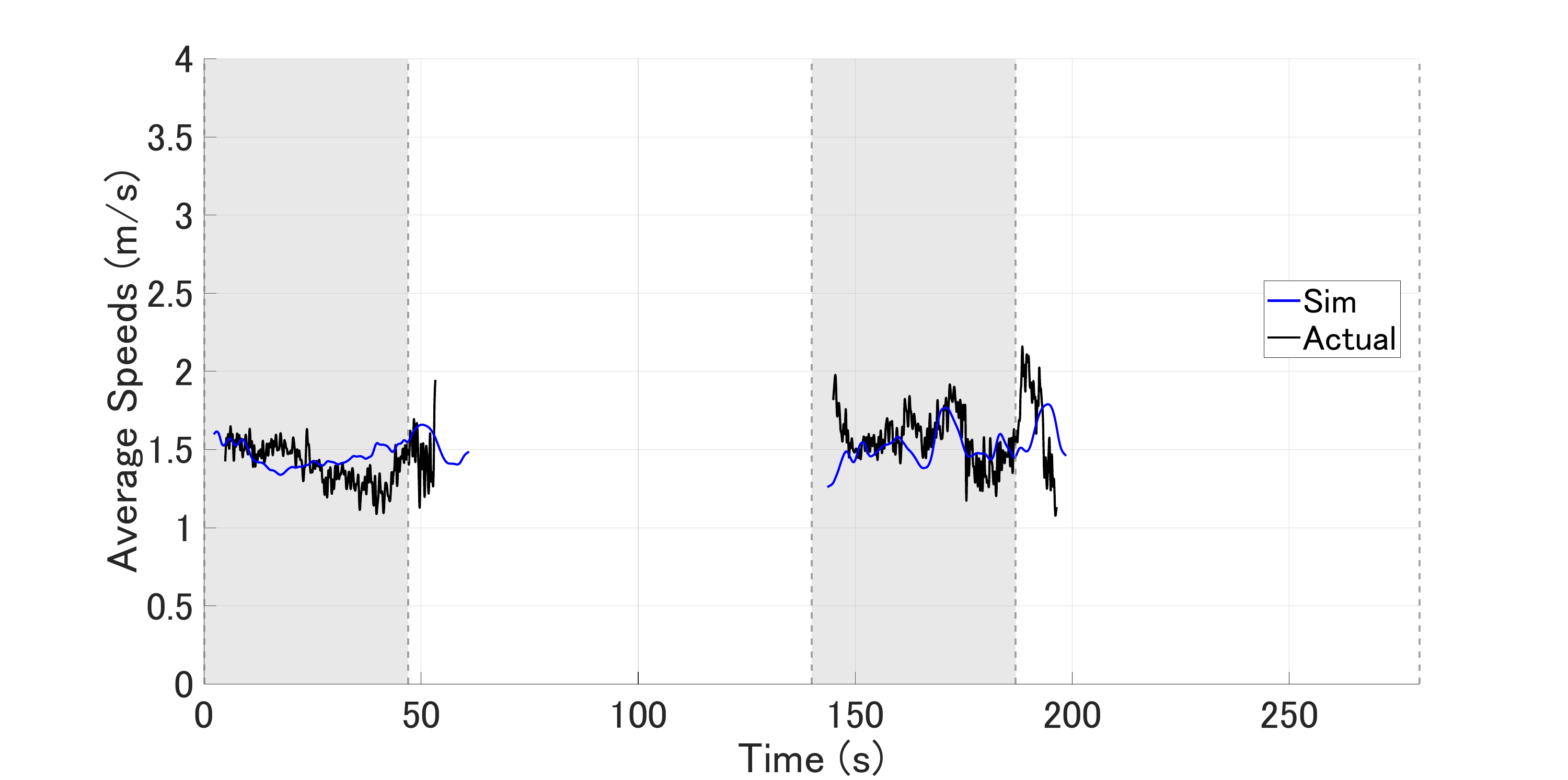}
        \subcaption{Speed of Flow 2}
        \label{fig:sub:WEEKDAY_SPEED_FLOW2}
      \end{minipage} 
    \end{tabular}
     \caption{(a) and (b) number of pedestrians within the crosswalk per frame for each flow during a weekday. (c) and (d) average speed per frame of pedestrians within the crosswalk during each flow on a weekday.}
  \label{fig:WEEKDAY_COMPARING}
  \end{figure*}
\subsection{Quantitative Validation}
\label{sect:sub:weekday_sim}
\subsubsection{Simulation Setting}
In this subsection, we quantitatively verify the validity of the simulations, comparing the number of pedestrians and average walking speed between simulation result and actual data, as these two quantities are important in evaluating flow rate and density of pedestrian flows.
For comparison, we replaced some of the pedestrians in the actual data with simulated agents and computed their trajectories.

In the following simulations, we simulated replaced pedestrians in Flows 1 and 2 for 280 s, corresponding to two cycles of the traffic light. 
A challenge in such a simulation setting is the fewer number of fully tracked pedestrians in the actual data.
As mentioned in Section \ref{sect:sub:data_acquisition}, the number of pedestrian trajectories that were fully tracked by SMILETrack was 407 in the 50-min data (183 for Flow 1 and 224 for Flow 2, as shown in Fig. \ref{fig:TRAJ_TRAIN_SOURCE}).
Extracting pedestrian trajectories for the simulated 280 s from the 50-min data, we obtained the 19 trajectories (6 for Flow 1 and 13 for Flow 2) shown in Fig. \ref{fig:WEEKDAY_REPLACED_TRAJECTORIES}.
This is approximately 8\% of the pedestrians in actual Flows 1 and 2.

Table \ref{tab:OBTAINED_TRAJECTORIES} shows the number of pedestrian trajectories obtained by SMILETrack or interpolation for Flows 1 and 2, the number estimated by manual counting, and the resulting percentage divided by estimated numbers.
We replaced 89 pedestrians from Flow 1 and 39 pedestrians from Flow 2 data with agents calculated by the trained model.
As described in Section \ref{sect:sub:data_acquisition}, few consistent pedestrian trajectories were obtained from start to destination. 
The number of consistent trajectories was 19, as shown in Fig. \ref{fig:WEEKDAY_REPLACED_TRAJECTORIES}\subref{fig:sub:TRAJ_TRAIN_SOURCE}, corresponding to part of the trajectories in Figs. \ref{fig:TRAJ_TRAIN_SOURCE}\subref{fig:sub:FLOW1_TRAIN_TRAJ} and \subref{fig:sub:FLOW2_TRAIN_TRAJ}, and the number of total replaced pedestrians was 128.
Therefore, we employed the trajectory interpolation method using the trained proposed model described in Appendix \ref{append:interpolation}.
Fig. \ref{fig:WEEKDAY_REPLACED_TRAJECTORIES}\subref{fig:sub:INTERPOLATED_TRAJECTORIES} shows 109 pedestrian trajectories obtained by interpolation.
During this period, approximately 160 pedestrians in Flow 1 and 60 pedestrians in Flow 2 were estimated. Therefore, Flow 1 was able to replace about 56 \% of the pedestrians and Flow 2 about 65 \% of the pedestrians.

\subsubsection{Result}
Fig. \ref{fig:SHIBUYA_NORMAL_SIMULATION} shows a snapshot of the simulations, in which red and blue points indicate the position of simulated pedestrians of Flows 1 and 2, respectively, and yellow points indicate the position of the other pedestrians included in the actual data. 
Simulated pedestrians appropriately behaved in a manner similar to the previous simulation, including walking, waiting for traffic light, and crossing at the crosswalk.
Also, simulated pedestrians collided with each other, as indicated by the yellow points on the other pedestrian data.

Figs. \ref{fig:WEEKDAY_COMPARING}\subref{fig:sub:WEEKDAY_NUMBER_FLOW1} and \subref{fig:sub:WEEKDAY_NUMBER_FLOW2} show the number of pedestrians within the crosswalk, and Figs. \ref{fig:WEEKDAY_COMPARING}\subref{fig:sub:WEEKDAY_SPEED_FLOW1} and \subref{fig:sub:WEEKDAY_SPEED_FLOW2} show the average walking speed. 
In each figure, the black solid line indicates the result obtained from the actual data whereas the red and blue solid lines indicate the results obtained from the simulation of Flows 1 and 2, respectively.
The gray background areas in the graphs indicate the periods when the traffic light was green.
Moreover, Table \ref{tab:NORMAL_RMSE} shows the root mean square error (RMSE) between the actual data and simulations for each metric.
\begin{itemize}
    \item {\bf Number of pedestrians}
    
    From Figs. \ref{fig:WEEKDAY_COMPARING}\subref{fig:sub:WEEKDAY_NUMBER_FLOW1} and \subref{fig:sub:WEEKDAY_NUMBER_FLOW2}, the overall shapes of the graphs are very similar between the actual data and simulations: the number of pedestrians increased after the traffic light turned green, reached a peak at similar timing, and then decreased as the traffic light turned red. The RMSE is 2.34 for Flow 1 and 1.88 for Flow 2.
    \item {\bf Average walking speed}
    
   From Figs. \ref{fig:WEEKDAY_COMPARING}\subref{fig:sub:WEEKDAY_SPEED_FLOW1} and \subref{fig:sub:WEEKDAY_SPEED_FLOW2},  the overall shapes of the average walking speeds are also very similar: both flows are high just after the traffic light turns green, and then decrease.
   Interestingly, average walking speed increases as the traffic light turns red possibly because in reality pedestrians walk faster, trying to finish crossing before the traffic light turns red.
   This result implies that the trained model reproduced such behavior.
   The RMSE was 0.17 m/s for Flow 1 and 0.20 m/s for Flow 2. 
\end{itemize}
These two results quantitatively validate that the trained model can reproduce not only individual pedestrian behavior but also crowd behavior of each pedestrian flow.

\begin{table}[t]
    \centering
    \caption{Results of RMSE comparison for the number of pedestrians and average walking speeds at the crosswalk for each flow in the simulations.}
    \begin{tabular}{l| c c}
        \toprule
        Flow & Number of pedestrians [persons] & Average speed [m/s] \\
        \midrule
        Flow 1 & 2.34 & 0.17 \\
        Flow 2 & 1.88 & 0.20 \\
        \bottomrule
    \end{tabular}
    \label{tab:NORMAL_RMSE}
\end{table}

\section{Discussion}
\label{sect:discussion}
\subsection{Tracking Error of Pedestrian Trajectories}
As mentioned in Section \ref{sect:sub:data_acquisition}, SMILETrack often failed to track the same person in the recorded video. 
As a result, fewer of the fully-tracked trajectories could be used for the learning or the simulation for comparative verification.
Synchronized multiple camera images can avoid tracking error due to occlusion or insufficient image resolution.

\subsection{Unnatural Behavior in Simulation}
When we increased the flow rate of the pedestrians and simulated their behaviors in a manner similar to Section \ref{sect:validation_of_simulation}, several pedestrians ignored the traffic light and entered the crosswalk even at red light.
This is because the correct answer rate for edges obtained through training was higher than 99 \%, yet not 100 \%.
In the proposed training model, the loss for such traffic light violations is equal to simple edge misclassification and treated on a par with position prediction errors.
Consequently, the model lacks the mechanism to explicitly suppress red-light-ignoring behavior.
One possible countermeasure is shift to a reinforcement learning framework, in which pedestrians crossing during a red traffic light are given a large negative reward so that the model learns to avoid such behavior.

\section{Conclusion}
\label{sect:conclusion}
A multi-scale model of pedestrian flows integrating global and local behaviors was proposed.
The proposed model employs a topological graph structure to represent global route selection behavior and an Attention mechanism in the decoder part to keep the global edge state transition and local position prediction consistent.
The obtained results are summarized as follows.
\begin{enumerate}
    \item 
    We recorded a video data of Shibuya scramble crossing using an 8K digital camera, obtaining a total 407 data pedestrian trajectories from a 50-min video: 183 for Flow 1 that started from the Shibuya station to the crosswalk, and 224 for Flow 2 in the opposite direction.
    The proposed model trained using this data 
    achieved an estimated pedestrian position error of 0.07 m and an accuracy more than 99\% for edge prediction regarding route selection, using 10\% of the data for verification.
    \item 
    Based on the trained model, we simulated pedestrian flows at the Shibuya scramble crossing by predicting the walking trajectories of virtual pedestrians.
    In the simulations, we observed natural behaviors of the state transitions such as stopping in front of the crosswalk at the red light, and starting walking after the traffic light turned green. 
    We also observed lane formation at the crosswalk despite adding a rule-based collision avoidance effect to the prediction by the trained model.
    In particular, some behaviors not explicitly considered in the model were observed in the data: some pedestrians started crossing just before the green light, or pedestrians in Flow 1 selected three branched routes around Node 4.
    These results qualitatively validate that the proposed model can reproduce natural pedestrian behaviors consistent with the data.    
    \item 
    As a quantitative evaluation, we compared the number of pedestrians and average walking speed within the crosswalk with those of the actual pedestrian trajectories.
    In comparisons, we simulated Flows 1 and 2, partially replacing corresponding data with those of virtual pedestrians.
    In both Flows 1 and 2, the simulations successfully reproduced profiles of the temporal variation of the number of pedestrians and average walking speed similar to the actual data.
    The RMSE for the number of pedestrians was 2.34 persons for Flow 1 and 1.88 persons for Flow 2, and the RMSE for average walking speed was 0.17 m/s for Flow 1 and 0.20 m/s for Flow 2.
    These results quantitatively validate the simulations based on the proposed model.
\end{enumerate}
One of the future works is a simulation of all pedestrian flows at the Shibuya scramble crossing, training the proposed model with more data.
The proposed model and simulations can be applied to the pedestrian flow VR simulator\cite{sakurai2023vr} which contributes to crowd navigation or evacuation in the event and disaster.

\section*{Acknowledgments}

The authors are grateful to K. Tomabechi and A. Saito for assistance with recording at Shibuya Scramble Crossing.

\refstepcounter{app}
\section*{Appendix \theapp \\ Conversion from pixel coordinates to 2D plane coordinates}
\addcontentsline{toc}{section}{Appendix \theapp: Conversion from pixel coordinates to 2D plane coordinates}
\label{append:pos3d_calculaion}
In \eqref{eq:picTo3D}, let $\bm{m}_i \ (i=1,2,3)$ denote a row vector in the matrix $\bm{M}$.
From the third row of \eqref{eq:picTo3D}, the scale parameter $s$ is obtained as follows:
\begin{align}
 s=\bm{m}_3^T\bm{x} .  
\end{align}
Substituting this into the first and second rows of \eqref{eq:picTo3D} and rearranging the equation yields:
\begin{align}
    \bm{C}\bm{x}&=\bm{0}, \quad
    \bm{C}:=\begin{bmatrix}
        u \\ v
    \end{bmatrix}\bm{m}_3^T-\begin{bmatrix}
        \bm{m}_1^T \\ \bm{m}_2^T
    \end{bmatrix} \label{eq:proj_rearranged} .
\end{align}
If we decompose the matrix $\bm{C}\in \mathbb{R}^{2 \times 4}$ into $\bm{A},\bm{B}\in \mathbb{R}^{2 \times 2}$ such that $\bm{C}=[\bm{A}\ \bm{B}]$ is satisfied, \eqref{eq:proj_rearranged} can be rewritten as follows:
\begin{align}
    \bm{A}\begin{bmatrix}
        X \\ Y
    \end{bmatrix} &= \bm{b}, \quad \bm{b}:=-\bm{B}\begin{bmatrix}
        Z \\ 1
    \end{bmatrix} .
\end{align}
If the value of $Z$ is known and $\bm{A}$ is regular, $X$ and $Y$ is obtained by $\begin{bmatrix}
    X \\ Y
\end{bmatrix}=\bm{A}^{-1}\bm{b}$.

In this study, the height of the bounding box was assumed to be 2 m because the average height of a Japanese male is 170 cm.
Then, we set $Z=1$ to select the center of the bounding box as the 3D position of each pedestrian.

\refstepcounter{app}
\section*{Appendix \theapp \\ Trajectory Interpolation using Prediction by Trained Pedestrian Flow Model}
\addcontentsline{toc}{section}{Appendix \theapp: Trajectory Interpolation using Prediction by Trained Pedestrian Flow Model}
\begin{figure}[tbp]
  \centering
  \begin{minipage}{0.48\linewidth}
    \centering
    \includegraphics[width=\linewidth]{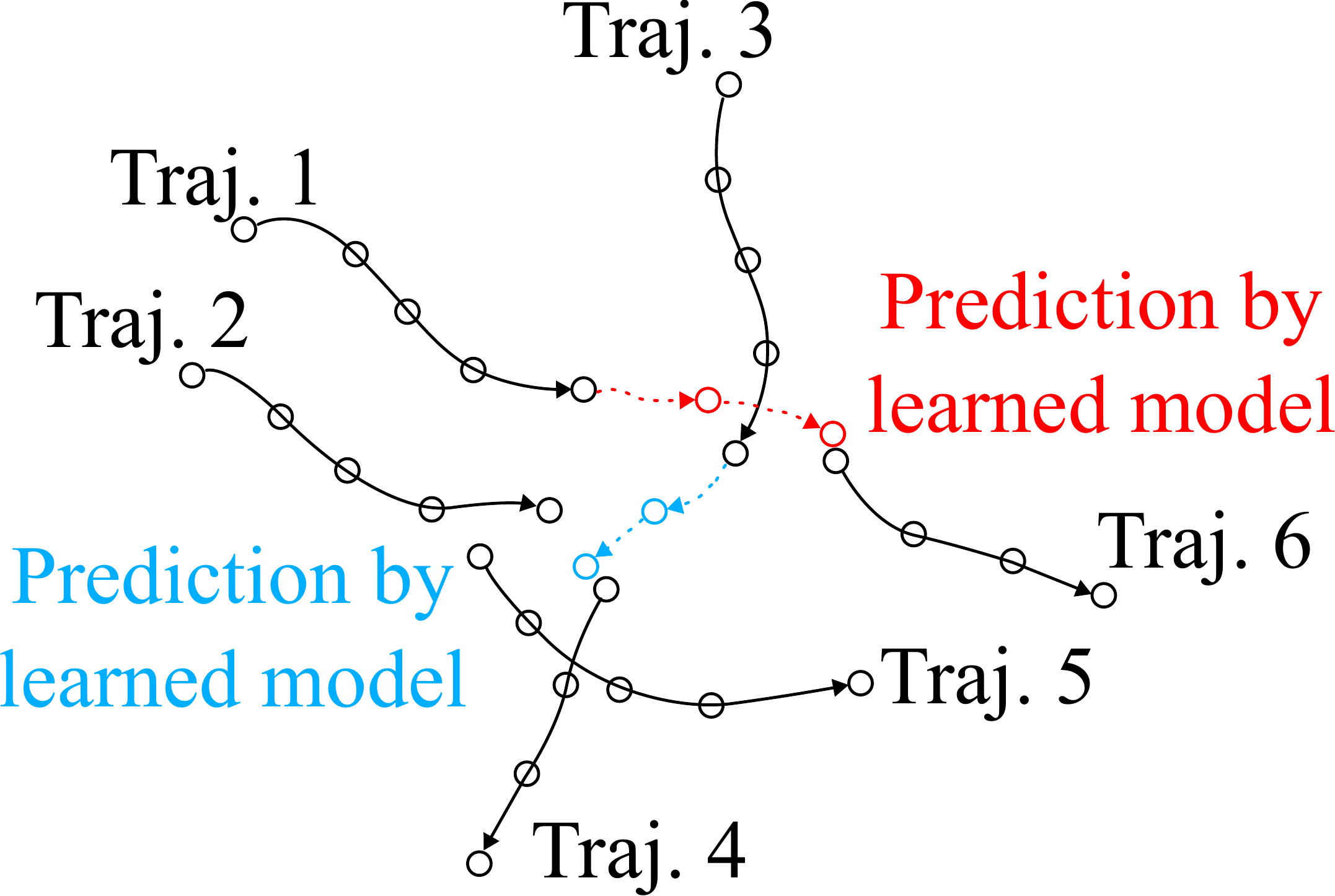}
    \subcaption{}
  \end{minipage}
  \hfill
  \begin{minipage}{0.48\linewidth}
    \centering
    \includegraphics[width=\linewidth]{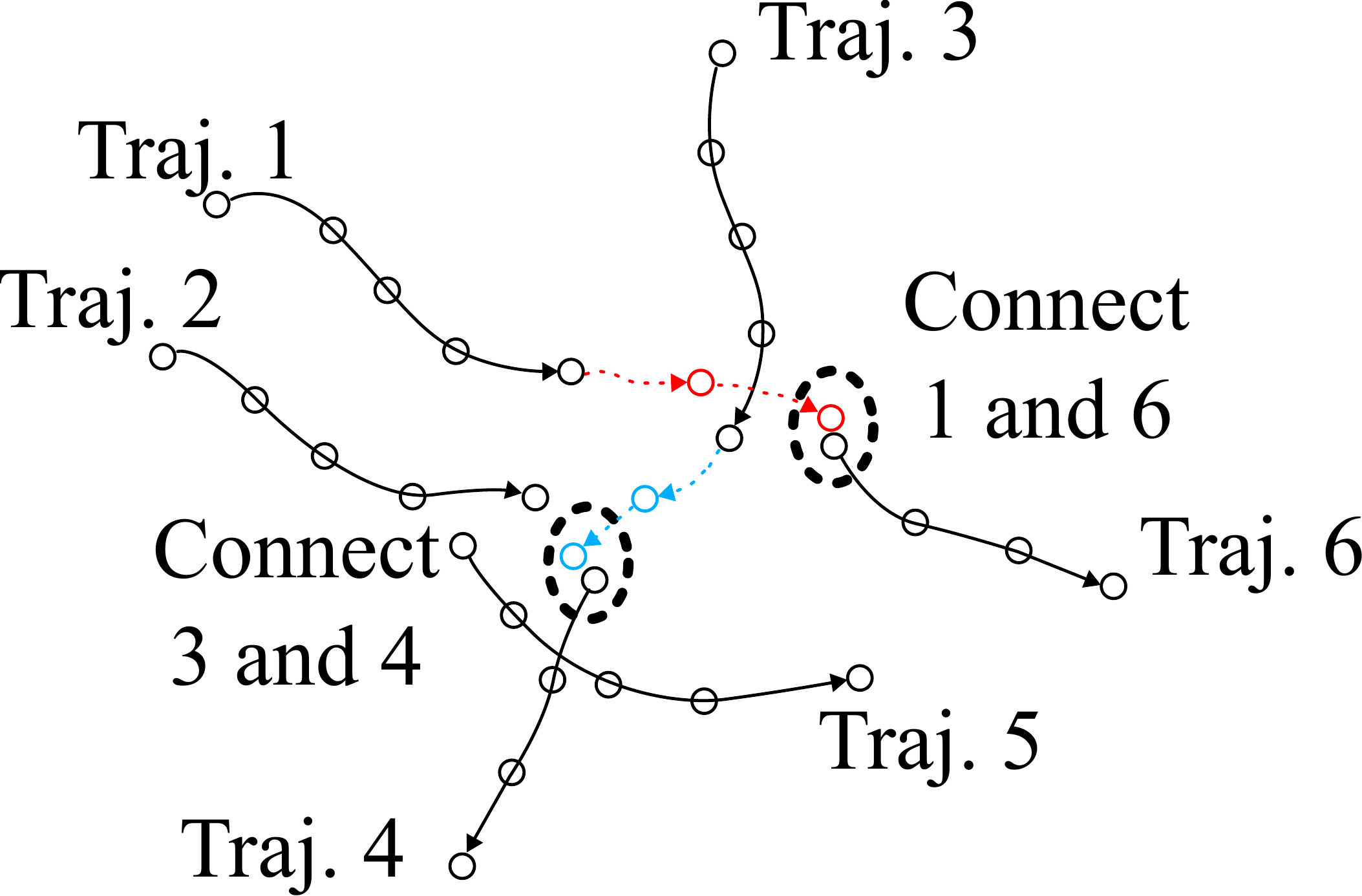}
    \subcaption{}
  \end{minipage}
  \caption{Schematic diagram of pedestrian trajectory interpolation. (a) Future trajectories predicted by trained models from each pedestrian trajectory. (b) Connected trajectories where the predicted point is close to the starting point.}
  \label{fig:INTERPOLATE_ALGO}
\end{figure}
\begin{figure}[t]
  \begin{center}
  \includegraphics[width=0.8\hsize]{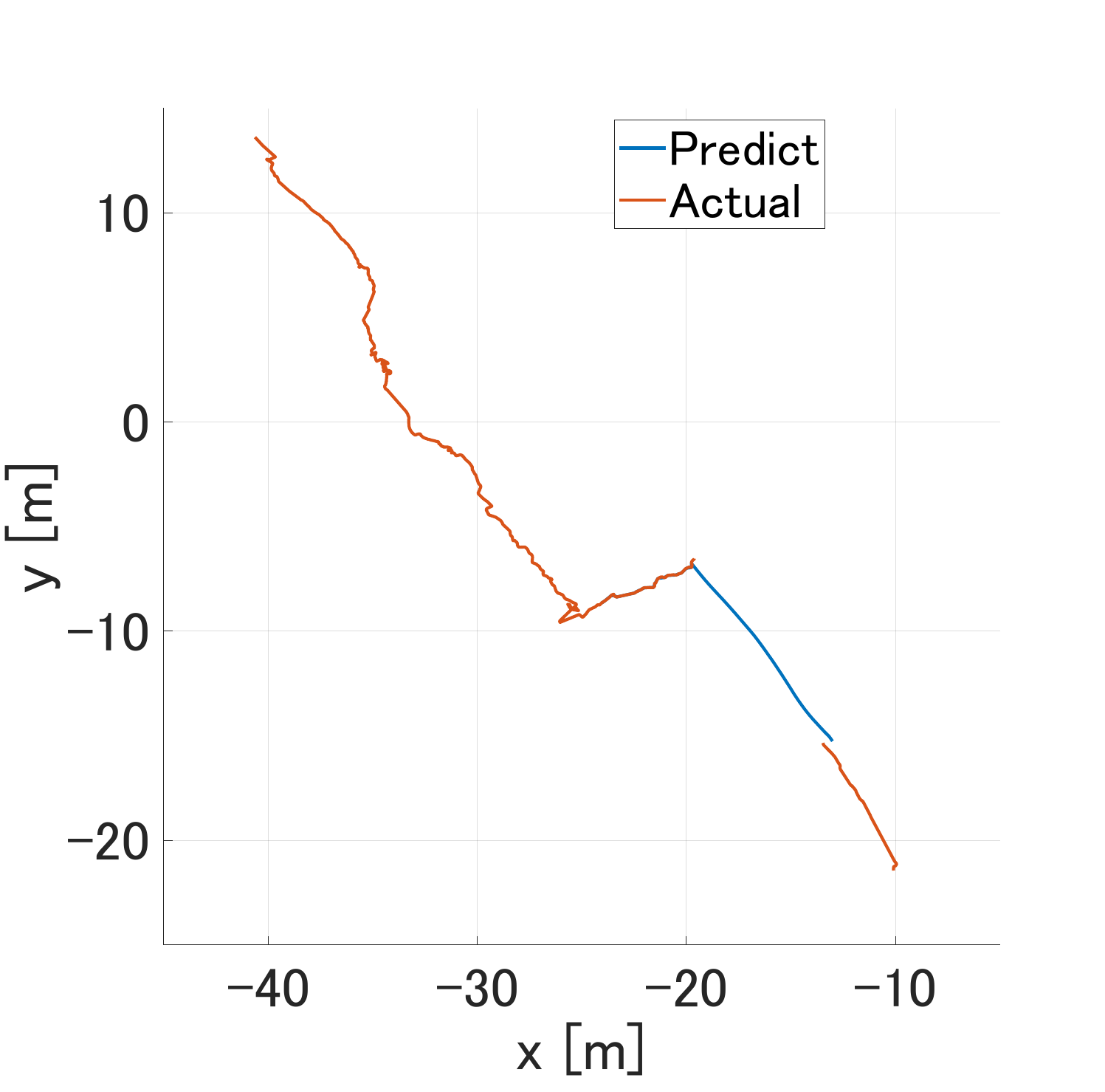}
  \caption{An example of pedestrian trajectory interpolation by prediction. Orange solid lines represent the actual pedestrian trajectories, whereas blue solid line represents the predicted trajectory. The prediction connects the actual pedestrian trajectories.}
  \label{fig:INTERPOLATION_EXAMPLE}
  \end{center}
\end{figure}
\label{append:interpolation}
Pedestrian trajectory data used for training consisted only of trajectories with time series of positions obtained for the same individual.
For example, in Flow 1, a pedestrian started from Node 1, which is the entrance of the station, and crossed the crosswalk to reach Node 4.
Such pedestrian trajectories account for approximately 20\% of the actual pedestrian count in each flow.
Although each pedestrian can be consistently tracked up to the point just before the crosswalk, tracking often failed at the point pedestrians waited at the crosswalk stop line or in the crosswalk. This is mainly due to insufficient lighting during recording, occlusions in dense situations such as during the blue traffic light waiting phase, and presence of pedestrians carrying umbrellas.
As a result, consistent pedestrian trajectories were few.
Hence, even the flow of pedestrians, which was the target of training in the simulation by the trained model, was reproduced, verifying the performance of the trained model as a group was difficult because an insufficient number of pedestrians could be simulated.

Therefore, utilizing the fact that target pedestrian flow belongs to the flow used for training, the trained model was employed to interpolate the missing segments.
Since the objective was to detect a sufficient number of pedestrians belonging to the flow used for training, missing observations of pedestrians from other flows were disregarded in this study.
A schematic diagram of the interpolation algorithm for the missing segments is shown in Fig. \ref{fig:INTERPOLATE_ALGO}.
When multiple trajectories are present, prediction begins from the endpoint of one trajectory, and if it can be connected to the starting point of another trajectory, the two are linked to interpolate the missing segment.
The detailed algorithm is presented in Algorithm \ref{alg:TRAJECTORY_CONNECTION}.
Here, $T$ denotes the set of pedestrian trajectories in the flow data: $F$ is the trained model; $N$ is the maximum number of prediction steps; and $\delta$ is the distance threshold used for trajectory connection.
\begin{algorithm}
    \caption{Trajectory Connection Procedure}
    \label{alg:TRAJECTORY_CONNECTION}
    \DontPrintSemicolon
    \KwIn{Set of trajectories $T$, trained flow model $F$, prediction limit $N$, distance threshold $\delta$}
    \KwOut{Connected trajectories $T'$}

    \BlankLine
    $T' \leftarrow [\;]$\;

    \BlankLine
    \ForEach{$\tau \in T$}{
      \If{$\tau$ is not part of the target flow}{
        \textbf{continue}\;
      }
      $p \leftarrow$ endpoint of $\tau$\;
      $\tau_{\text{conn}} \leftarrow \tau$\;
      \For{$i=1$ \KwTo $N$}{
        $p_{\text{pred}} \leftarrow F.\text{predict}(p)$\;
        \If{\text{\upshape there exists }$\tau' \in T$\\\text{\upshape  such that }$dist(p_{\text{pred}}, $\text{\upshape startpoint of }$\tau') < \delta$}{
          connect $\tau_{\text{conn}}$ and $\tau'$\;
          $p \leftarrow$ endpoint of $\tau'$\;
          $i \leftarrow 1$\;
          \textbf{continue}\;
        }
      }
      add $\tau_{\text{conn}}$ to $T'$\;
    }
    \BlankLine
    \Return $T'$\;
\end{algorithm}

Fig. \ref{fig:INTERPOLATION_EXAMPLE} shows an example of interpolating actual pedestrian trajectories using the proposed prediction-based method.
Two trajectories that were originally separated in space are successfully connected through prediction, allowing them to be treated as valid trajectories for the pedestrians to be simulated.
Fig. \ref{fig:WEEKDAY_REPLACED_TRAJECTORIES}\subref{fig:sub:INTERPOLATED_TRAJECTORIES} presents the trajectories that were replaced in the weekday reproduction simulation.


\bibliographystyle{IEEEtran}
\bibliography{bib/examples/IEEEabrv, bib/my_bib}

\vfill

\end{document}